\documentclass[11pt, a4paper]{article}
\pdfoutput=1

\usepackage{jheppub}
\usepackage{graphicx, bm, color, amssymb, amsmath,slashed}
\usepackage{float}
\usepackage[caption=false]{subfig}
\usepackage{tabularx}

\newcommand{\tr}[1]{\textrm{#1}}

\newcolumntype{L}[1]{>{\raggedright\arraybackslash}p{#1}}
\newcolumntype{C}[1]{>{\centering\arraybackslash}p{#1}}
\newcolumntype{R}[1]{>{\raggedleft\arraybackslash}p{#1}}

\begin{document}
\title{Cornering dimension-6 $HVV$ interactions
       at high energy LHC: the role of event ratios}
\author{Shankha Banerjee$^1$, Tanumoy Mandal$^1$, Bruce Mellado$^2$, Biswarup Mukhopadhyaya$^1$}
\affiliation[1]{Regional Centre for Accelerator-based Particle Physics,
Harish-Chandra Research Institute, \\ 
Chhatnag Road, Jhusi, Allahabad  211019, India}
\affiliation[2]{School of Physics, University of the Witwatersrand, Wits 2050, South Africa}
\emailAdd{shankha@hri.res.in}
\emailAdd{tanumoymandal@hri.res.in}
\emailAdd{bruce.mellado@wits.ac.za}
\emailAdd{biswarup@hri.res.in}
\preprint{HRI-RECAPP-2015-008, WITS-MITP-006}
\abstract{We suggest a way of improving the probes on dimension-6
  CP-conserving $HVV$ interactions ($V$ = $W$, $Z$, $\gamma$), from the LHC
  data on the Higgs boson to be available in the 14 TeV run with an integrated luminosity of $3000$ fb$^{-1}$.  We find
  that the ratios of total rates in different channels can be quite
  useful in this respect.  This includes ratios of event rates in (a)
  different final states for the Higgs produced by the same production mechanism,
  and (b) the same final state from two different production
  modes. While most theoretical uncertainties cancel in the former,
  the latter helps in the case of those operators which shift the
  numerator and denominator in opposite directions. Our analysis,
  incorporating theoretical, systematic and statistical uncertain, leads
  to projected limits that are better than the strongest ones obtained so
  far from precision electroweak as well as LHC Higgs data.
   Moreover, values of the coefficients of the dimension-6 operators, which are
  allowed in disjoint intervals, can have their ranges narrowed down
  substantially in our approach. }

\maketitle
\section{Introduction}
\label{sec:intro}

The ATLAS and CMS experiments at the Large Hadron Collider (LHC) have discovered a neutral
spinless particle that closely matches the description of the Higgs
boson~\cite{Aad:2012tfa,Chatrchyan:2012ufa} which is responsible for 
masses of elementary particles, according
to the standard model (SM) of electroweak interactions. While this ties the final 
knot on the framework embodied in the SM, there are many reasons to believe that there 
is more fundamental physics at higher energies. The reason for such
expectation can be traced to many issues, including the unexplained 
replication of fermion families, the source of dark matter in the universe,
and the problems of naturalness and vacuum stability involving the Higgs 
boson itself. The Large Hadron Collider (LHC) has not revealed any direct signature of
new physics so far. However, one is led to suspect that such physics should  
affect the interaction Lagrangian of the Higgs boson. This generates, for example,
effective operators of dimension-6 contributing to $HVV$ interactions,
with $V=W,Z,\gamma$. Probing such effective couplings for the recently
discovered scalar is therefore tantamount to opening a gateway to fundamental
physics just beyond our present reach.

Such `effective' interaction terms better be $SU(2) \times U(1)$
invariant if they arise from physics above the electroweak
scale. Constraints on such terms have already been studied, using
precision electroweak data as well as global fits of the current Higgs
data~\cite{Masso:2012eq,Corbett:2012ja,Falkowski:2013dza,Corbett:2013pja,Dumont:2013wma,Banerjee:2012xc,Gainer:2013rxa,
Corbett:2013hia,Elias-Miro:2013mua,Pomarol:2013zra,Einhorn:2013tja,Banerjee:2013apa,Willenbrock:2014bja,
Ellis:2014dva,Belusca-Maito:2014dpa,Gupta:2014rxa,Masso:2014xra,Biekoetter:2014jwa,Englert:2014cva,
Ellis:2014jta,Edezhath:2015lga,Gorbahn:2015gxa,Han:2004az,Ciuchini:2013pca,Blas:2013ana,Chen:2013kfa,
Alonso:2013hga,Englert:2014uua,Trott:2014dma,Falkowski:2014tna,Henning:2014wua,deBlas:2014mba,
Berthier:2015oma,Efrati:2015eaa,Bhattacherjee:2015xra}. 
Recently, CMS has published an exhaustive study on anomalous $HVV$ couplings~\cite{Khachatryan:2014kca}.
Many studies have considered anomalous Higgs
couplings in context of future lepton colliders~\cite{Amar:2014fpa,Kumar:2014zra,Craig:2014una,Beneke:2014vqa,Kumar:2015eea,Ren:2015uka}.
The general conclusion, based on analyses of the 8 TeV data, is that several (though not all) of the gauge invariant, dimension-6 $HVV$ 
terms have been quite strongly constrained by the EW precision and LHC data (as discussed in section~\ref{sec:ratio})~\cite{Masso:2012eq,Corbett:2012ja,Falkowski:2013dza,Corbett:2013pja,Banerjee:2012xc,Dumont:2013wma,Gainer:2013rxa,
Corbett:2013hia,Elias-Miro:2013mua,Pomarol:2013zra,Einhorn:2013tja,Banerjee:2013apa,Willenbrock:2014bja,
Ellis:2014dva,Belusca-Maito:2014dpa,Gupta:2014rxa,Masso:2014xra,Biekoetter:2014jwa,Englert:2014cva,
Ellis:2014jta,Edezhath:2015lga,Gorbahn:2015gxa,Han:2004az,Ciuchini:2013pca,Blas:2013ana,Chen:2013kfa,
Alonso:2013hga,Englert:2014uua,Trott:2014dma,Falkowski:2014tna,Henning:2014wua,deBlas:2014mba,
Berthier:2015oma,Efrati:2015eaa,Bhattacherjee:2015xra}. 
It still remains to be seen whether  such small coefficients can be discerned with some ingeniously constructed kinematic 
distributions. Some work has nonetheless been done to study such
distributions~\cite{Plehn:2001nj,Bernaciak:2012nh,Bernaciak:2013dwa,Biswal:2012mp,Djouadi:2013yb}, in terms of either the gauge invariant operators
themselves or the structures finally ensuing from them. At the same time,
it is of interest to see if meaningful constraints do arise from
the study of total rates at the LHC. The essence of any probe of these anomalous
couplings, however, lies in pinning them down to much smaller values 
using the 14 TeV runs, as common sense suggests the manifestation,
if any, of new physics  through Higher Dimensional Operators (HDO's) with small coefficients only.

We show here that the
relative rates of  events of different kinds in the Higgs data
can allow us to probe such effective interactions to levels of smallness 
not deemed testable otherwise~\cite{Djouadi:2012rh,Djouadi:2013qya}. This happens through (a) the cancellation of
theoretical uncertainties, and (b) the fact that some ratios have the numerators
and denominators shifting in opposite directions, driven by 
the additional interactions. Thus the cherished scheme of finding traces 
of new physics in Higgs phenomenology can be buttressed with one more brick.

We organise our paper as follows: we summarise the relevant gauge invariant operators
and the interaction terms in Sec.~\ref{sec:HDO}. In Sec.~\ref{sec:ratio}, 
we introduce three ratios of cross-sections as our observables. The results of our analysis are explained
in Sec.~\ref{results}. We summarise and conclude in Sec.~\ref{summary}.

\section{Higher dimensional operators}
\label{sec:HDO}

In order to see any possible deviations from the SM in the Higgs
sector, we will follow the effective field theory (EFT) framework. We
consider $SU(2)_L \times U(1)_Y$ invariant operators of dimension up to
6, which affect Higgs couplings to itself and/or a pair of electroweak
vector bosons.  While a full list of such operators are found in
~\cite{Buchmuller:1985jz,Hagiwara:1993qt,GonzalezGarcia:1999fq,Grzadkowski:2010es}, we have concentrated here
on dimension-6 CP-conserving operators which affect Higgs phenomenology. They include:

\begin{itemize}

\item
Operators which contain the Higgs doublet $\Phi$ and its derivatives:
\begin{equation}
\mathcal{O}_{\Phi,1} = (D_{\mu}\Phi)^{\dagger}\Phi\Phi^{\dagger}(D^{\mu}\Phi);~~~
\mathcal{O}_{\Phi,2} = \frac{1}{2}\partial_{\mu}(\Phi^{\dagger}\Phi)\partial^{\mu}(\Phi^{\dagger}\Phi);~~~
\mathcal{O}_{\Phi,3} = \frac{1}{3}(\Phi^{\dagger}\Phi)^{3}
\end{equation}
\item
Those containing $\Phi$ (or its derivatives) and the bosonic field strengths :
\begin{equation}
\mathcal{O}_{GG} = \Phi^{\dagger}\Phi G_{\mu\nu}^a G^{a\,\mu\nu};~~~
\mathcal{O}_{BW} = \Phi^{\dagger}\hat{B}_{\mu \nu} \hat{W}^{\mu \nu} \Phi;~~~
\mathcal{O}_{WW} = \Phi^{\dagger}\hat{W}_{\mu \nu} \hat{W}^{\mu \nu} \Phi \nonumber
\end{equation}
\begin{equation}
\mathcal{O}_{W}  = (D_{\mu}\Phi)^{\dagger} \hat{W}^{\mu \nu} (D_\nu \Phi);~~~
\mathcal{O}_{BB} = \Phi^{\dagger}\hat{B}_{\mu \nu} \hat{B}^{\mu \nu} \Phi;~~~
\mathcal{O}_{B}  = (D_{\mu}\Phi)^{\dagger} \hat{B}^{\mu \nu} (D_\nu \Phi),
\end{equation}
\end{itemize}
where 
\begin{equation}
 \hat{W}^{\mu \nu}=i\,\frac{g}{2} \sigma_{a}W^{a \; \mu \nu};~~~
 \hat{B}^{\mu \nu}=i\,\frac{g}{2}' B^{\mu \nu} \nonumber
\end{equation}
and $g$, $g'$ are respectively the $\tr{SU}(2)_\tr{L}$ and $\tr{U}(1)_\tr{Y}$ gauge couplings. $W^a_{\mu
  \nu} = \partial_{\mu}W^a_{\nu}-\partial_{\nu}W^a_{\mu} - g
\epsilon^{abc}W^b_{\mu} W^c_{\nu}$, $B_{\mu \nu} =
\partial_{\mu}B_{\nu}-\partial_{\nu}B_{\mu}$ and $G^a_{\mu \nu} =
\partial_{\mu}G^a_{\nu}-\partial_{\nu}G^a_{\mu} - g_s f^{abc}G^b_{\mu}
G^c_{\nu}$. The covariant
derivative of $\Phi$ is given as $D_{\mu}\Phi=(\partial_{\mu}+\frac{i}{2}g'
B_{\mu} + i g \frac{\sigma_a}{2}W^a_{\mu})\Phi$.
The Lagrangian in the presence of the above operators can be generally
expressed as:
\begin{equation} 
\mathcal{L} \supset \kappa\left(\frac{2 m_W^2}{v} H W_{\mu}^+ W^{\mu
  -}+\frac{ m_Z^2}{v} H Z_{\mu} Z^{\mu } \right) + \sum_{i}
\frac{f_{i}}{\Lambda^2}\mathcal{O}_{i}
\label{Lag},
\end{equation}
where in addition to the dimension-6 (D6) operators, we also allow for the SM-like
$HWW$ and $HZZ$ couplings to be scaled by a factor $\kappa$. 
While $\kappa \neq 1$ is indicative of certain kinds of 
new physics, we are specially interested in this study in the new observable features 
associated with the HDOs. Therefore, we have set $\kappa = 1$ for simplicity.\footnote{Possible 
constraints on the departure of $\kappa$ from unity have been obtained in the literature from 
global fits of the Higgs data (See for example~\cite{Masso:2012eq,Corbett:2012ja,Falkowski:2013dza,Corbett:2013pja,Dumont:2013wma,Banerjee:2012xc,Gainer:2013rxa,
Corbett:2013hia,Elias-Miro:2013mua,Pomarol:2013zra,Einhorn:2013tja,Banerjee:2013apa,Willenbrock:2014bja,
Ellis:2014dva,Belusca-Maito:2014dpa,Gupta:2014rxa,Masso:2014xra,Biekoetter:2014jwa,Englert:2014cva,
Ellis:2014jta,Edezhath:2015lga,Gorbahn:2015gxa,Han:2004az,Ciuchini:2013pca,Blas:2013ana,Chen:2013kfa,
Alonso:2013hga,Englert:2014uua,Trott:2014dma,Falkowski:2014tna,Henning:2014wua,deBlas:2014mba,
Berthier:2015oma,Efrati:2015eaa})}

No operator of the form $\mathcal{O}_{GG}$ is assumed to exist since
we are presently concerned with Higgs interactions with a pair
of electroweak vector bosons only. The operator $\mathcal{O}_{\Phi,1}$  is severely constrained by
the $T$-parameter (or equivalently the $\rho$ parameter), as  it alters the  $HZZ$ and
$HWW$ couplings by unequal multiplicative factors. As far as
$HZZ$ and
$HWW$ interactions are concerned, $\mathcal{O}_{\Phi,2}$
only scales the standard model-like couplings ($\kappa$), without bringing in any new
Lorentz structure. This amounts to a renormalization of the Higgs field. It also alters the Higgs self-coupling, something that
is the sole consequence of  $\mathcal{O}_{\Phi,3}$ as well.

In view of the above, we focus on the four operators $\mathcal{O}_{WW}$, $\mathcal{O}_{BB}$,
$\mathcal{O}_W$ and $\mathcal{O}_B$.  We do not include the operator
$\mathcal{O}_{BW} = \Phi^{\dagger}\hat{B}_{\mu \nu} \hat{W}^{\mu \nu}
\Phi$ in the present analysis, because it mixes the
$Z$ and $\gamma$ fields at the tree level, violates custodial symmetry
(by contributing only to the $Z$-boson mass) and is, therefore, highly constrained
by the $S$ and $T$-parameters at the tree level~\cite{Corbett:2012ja}.
The effective interactions that finally emerge and affect the Higgs sector are 
\begin{align}
\label{eq:lagHVV}
\mathcal{L}_{eff} &= 
g_{HWW}^{(1)}~(W_{\mu\nu}^{+}W^{-\mu}\partial^{\nu}H + h.c.) +
g_{HWW}^{(2)}~HW_{\mu\nu}^{+}W^{-\mu\nu} \nonumber \\
&+ g_{HZZ}^{(1)}~Z_{\mu\nu}Z^{\mu}\partial^{\nu}H +
g_{HZZ}^{(2)}~HZ_{\mu\nu}Z^{\mu\nu} \nonumber \\
&+ g_{HZ\gamma}^{(1)}~A_{\mu\nu}Z^{\mu}\partial^{\nu}H +
g_{HZ\gamma}^{(2)}~HA_{\mu\nu}Z^{\mu\nu}+g_{H\gamma\gamma}H A_{\mu \nu} A^{\mu \nu},
\end{align}
where
\begin{align}
\label{eq:lagHVVcoeff}
g^{(1)}_{HWW}&=\left(\frac{g M_W}{\Lambda^2}\right) \frac{f_W}{2};~~~ g^{(2)}_{HWW}=-\left(\frac{g M_W}{\Lambda^2}\right)f_{WW} \nonumber \\
g^{(1)}_{HZZ}&=\left(\frac{g M_W}{\Lambda^2}\right) \frac{c^2 f_W + s^2 f_B}{2 c^2};~~~g^{(2)}_{HZZ}=-\left(\frac{g M_W}{\Lambda^2}\right) \frac{s^4 f_{BB} + c^4 f_{WW}}{2 c^2} \nonumber \\
g^{(1)}_{HZ\gamma}&=\left(\frac{g M_W}{\Lambda^2}\right)\frac{s(f_W-f_B)}{2 c};~~~g^{(2)}_{HZ\gamma}=\left(\frac{g M_W}{\Lambda^2}\right)\frac{s(s^2 f_{BB}-c^2 f_{WW})}{c} \nonumber \\
g_{H\gamma\gamma}&=-\left(\frac{g M_W}{\Lambda^2}\right)\frac{s^2(f_{BB}+f_{WW})}{2}
\end{align} with $s\,(c)$ being the sine (cosine) of the Weinberg angle.
Besides, the operators $\mathcal{O}_W$, $\mathcal{O}_B$ and $\mathcal{O}_{WWW}$ also contribute to the anomalous triple gauge 
boson interactions which can be summarised as 
\begin{align}
\label{eq:lagWWV}
\mathcal{L}_{WWV}=-i g_{WWV}\left\{g_1^V\left(W_{\mu\nu}^+W^{-\mu}V^{\nu}-W_{\mu}^+V_{\nu}W^{-\mu \nu}\right)+\kappa_V W_{\mu}^+W_{\nu}^-V^{\mu \nu} + \frac{\lambda_V}{M_W^2}W_{\mu \nu}^+ W^{-\nu \rho} V_{\rho}^{\mu}\right\},
\end{align}
where $g_{WW\gamma}=g \, s$, $g_{WWZ} = g \, c$, $\kappa_V=1+\Delta\kappa_V$ and $g_1^Z=1+\Delta g_1^Z$ with 
\begin{align}
\label{eq:lagWWVcoeff}
\Delta \kappa_{\gamma}&=\frac{M_W^2}{2
  \Lambda^2}\left(f_W+f_B\right);~~~\lambda_{\gamma}=\lambda_Z=\frac{3g^2M_W^2}{2\Lambda^2}
f_{WWW} \nonumber \\ \Delta g_1^Z&=\frac{M_W^2}{2 c^2 \Lambda^2}
f_W;~~~\Delta \kappa_Z=\frac{M_W^2}{2 c^2 \Lambda^2}\left(c^2 f_W -
s^2 f_B\right)
\end{align}

The already existing limits on the various operators discussed above
are found in numerous
references~\cite{Corbett:2013pja,Falkowski:2013dza,Corbett:2013hia,Masso:2012eq,Corbett:2012ja}. Even
within their current limits, some of the operators are found to modify the
efficiencies of the various kinetic cuts~\cite{Gainer:2013rxa,Banerjee:2013apa}. The question we address
in the rest of the paper is : can these limits be improved in the next run(s)
through  careful measurement of the ratios of total rates in different channels?
As we shall see below, the answer is in the affirmative.

\section{Ratios of cross-sections as chosen observables}
\label{sec:ratio}

\begin{figure*}
\centering
\subfloat[]{\includegraphics[width=4.8cm,height=4.8cm]{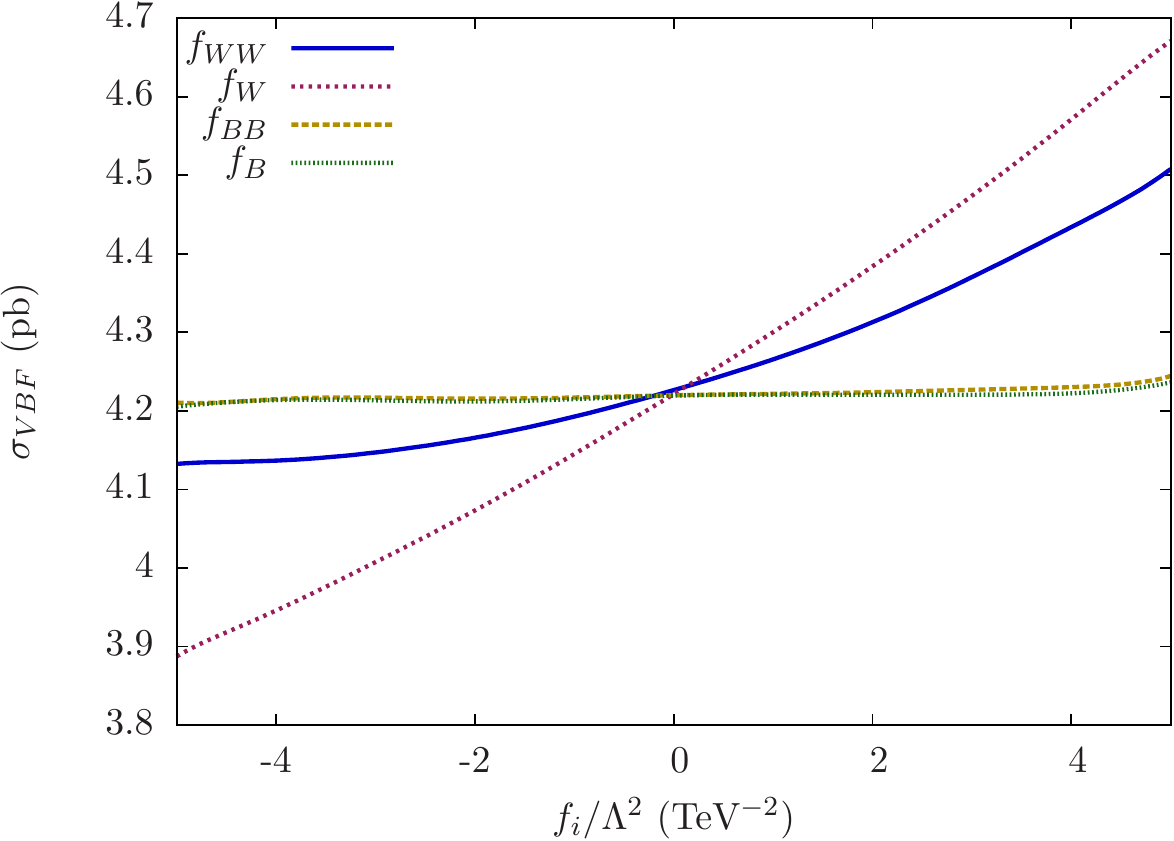}\label{fig:VBF}}~~
\subfloat[]{\includegraphics[width=4.8cm,height=4.8cm]{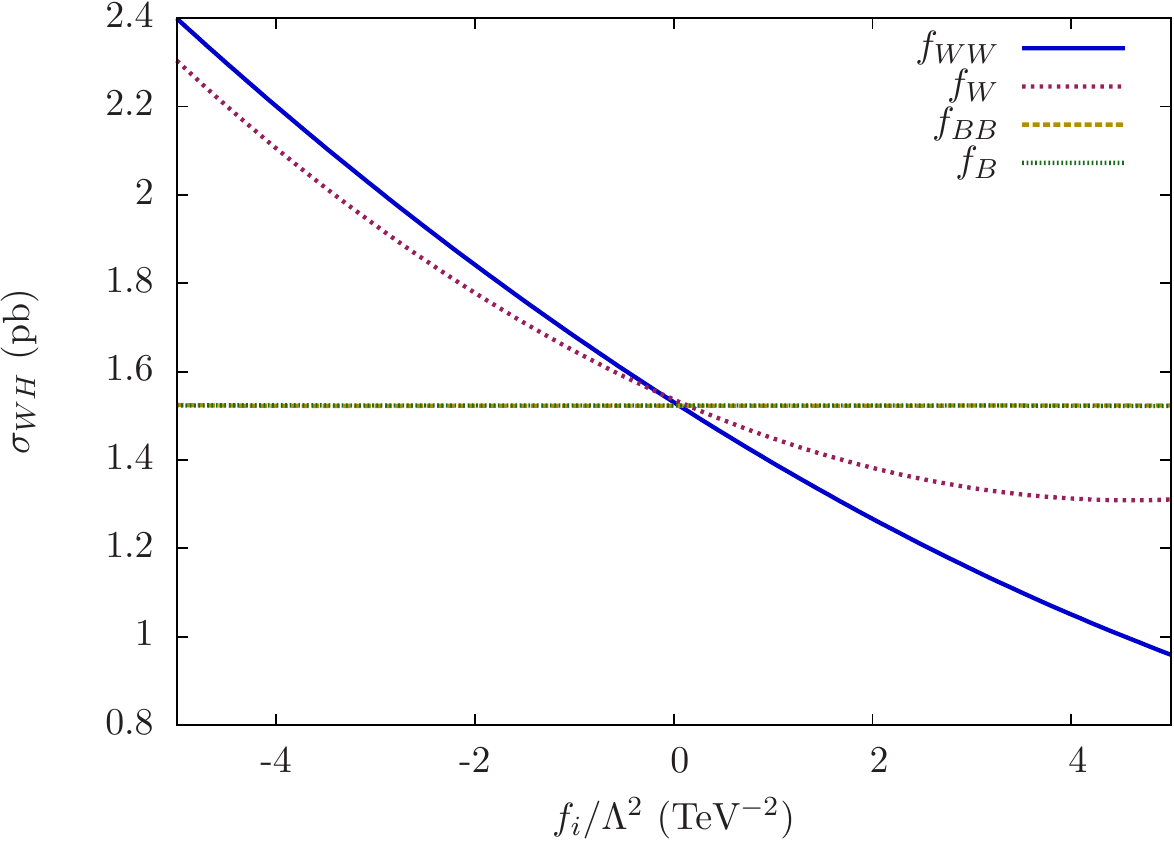}\label{fig:WH}}~~
\subfloat[]{\includegraphics[width=4.8cm,height=4.8cm]{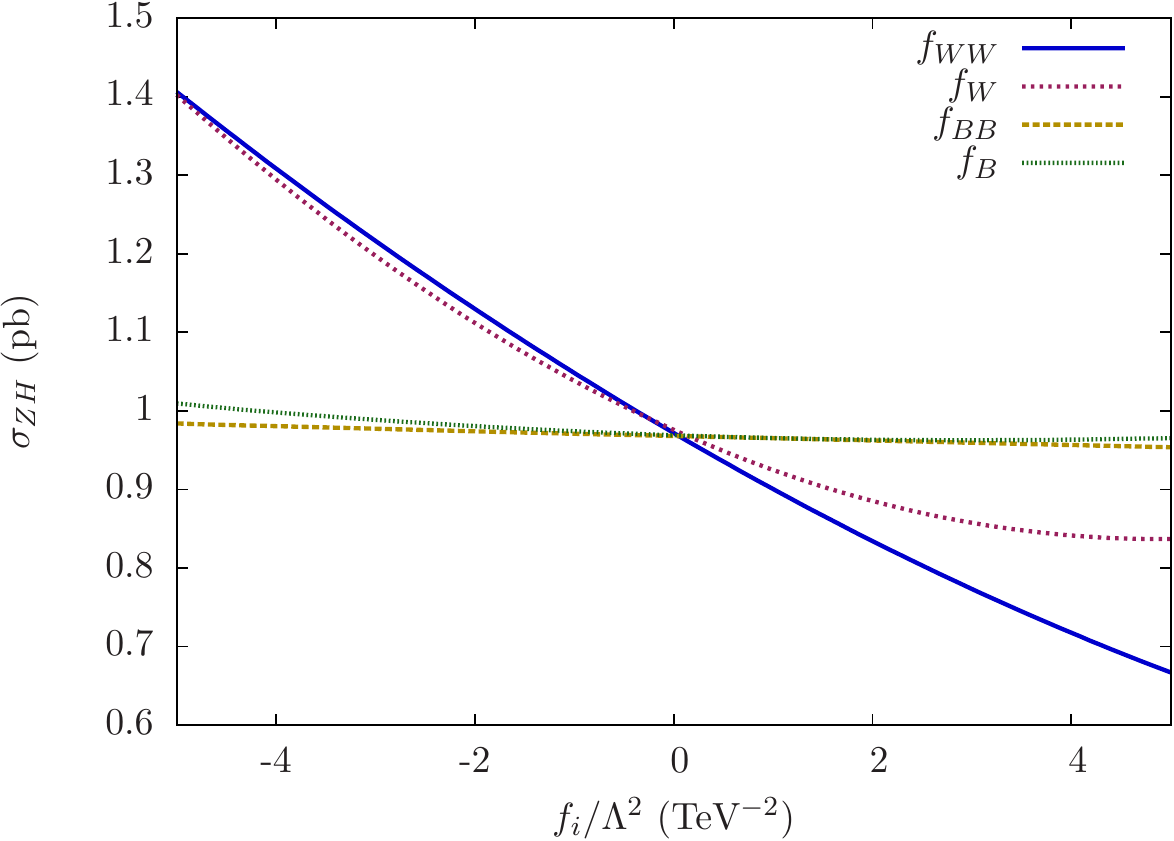}\label{fig:ZH}}
\caption{Higgs production cross-sections for the $VBF$ and $VH$
  channels in presence of HDOs at 14 TeV. Here the operators are varied one at a time.}
\label{fig:Hprod}
\end{figure*}

\begin{figure*}
\centering
\subfloat[]{\includegraphics[width=6cm,height=6cm]{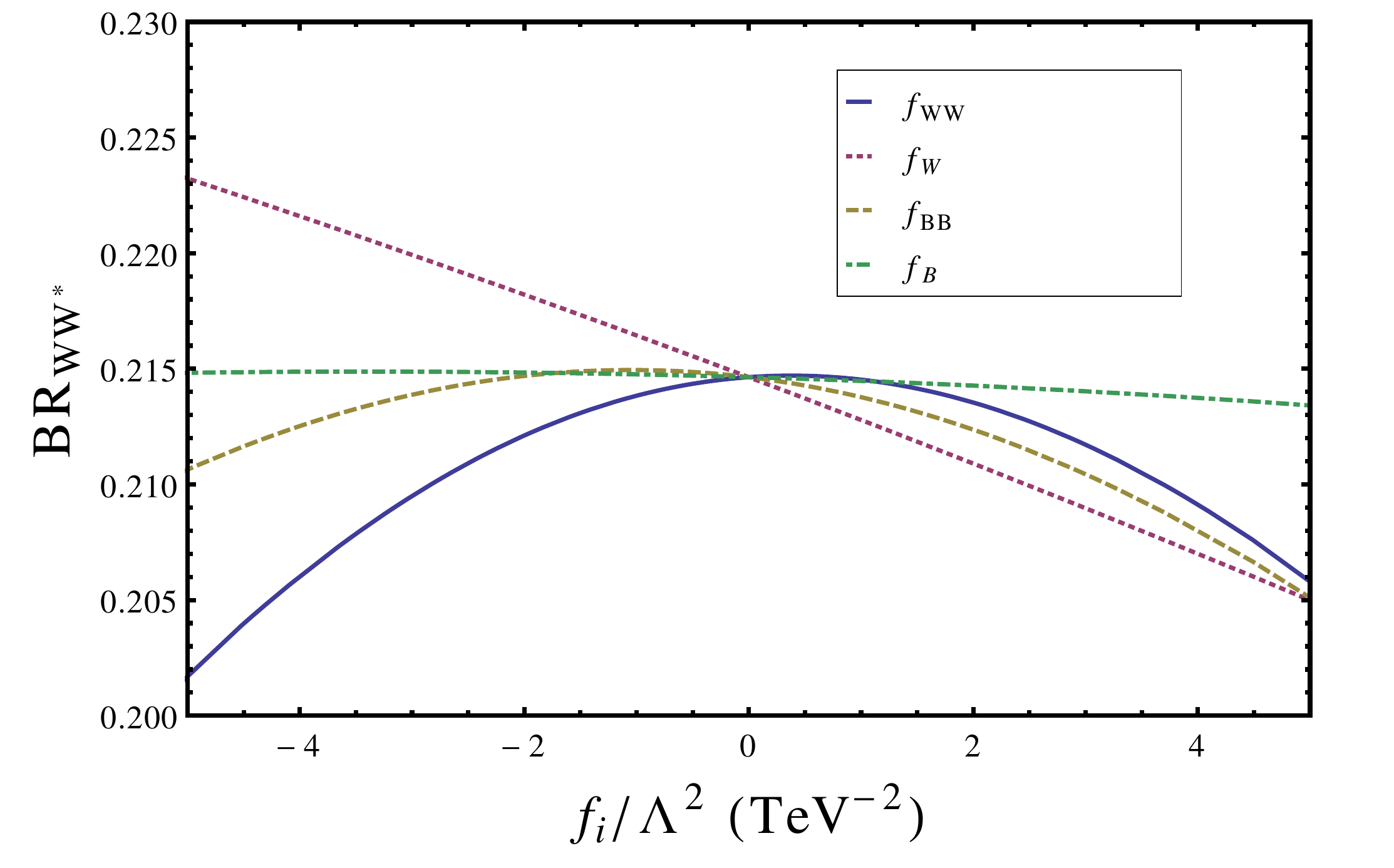}\label{fig:BRww}}~~~
\subfloat[]{\includegraphics[width=6cm,height=6cm]{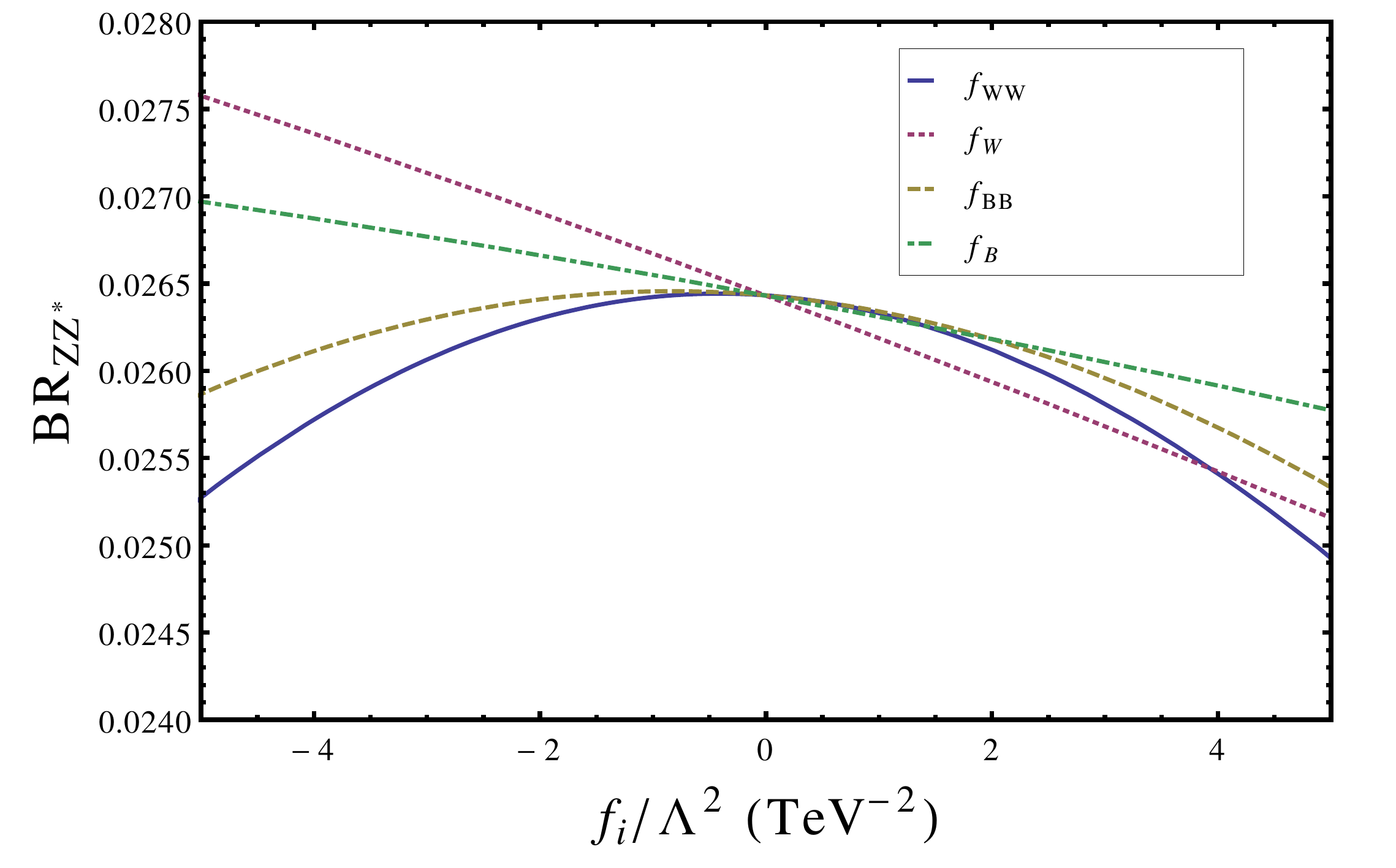}\label{fig:BRzz}}\\
\subfloat[]{\includegraphics[width=6cm,height=6cm]{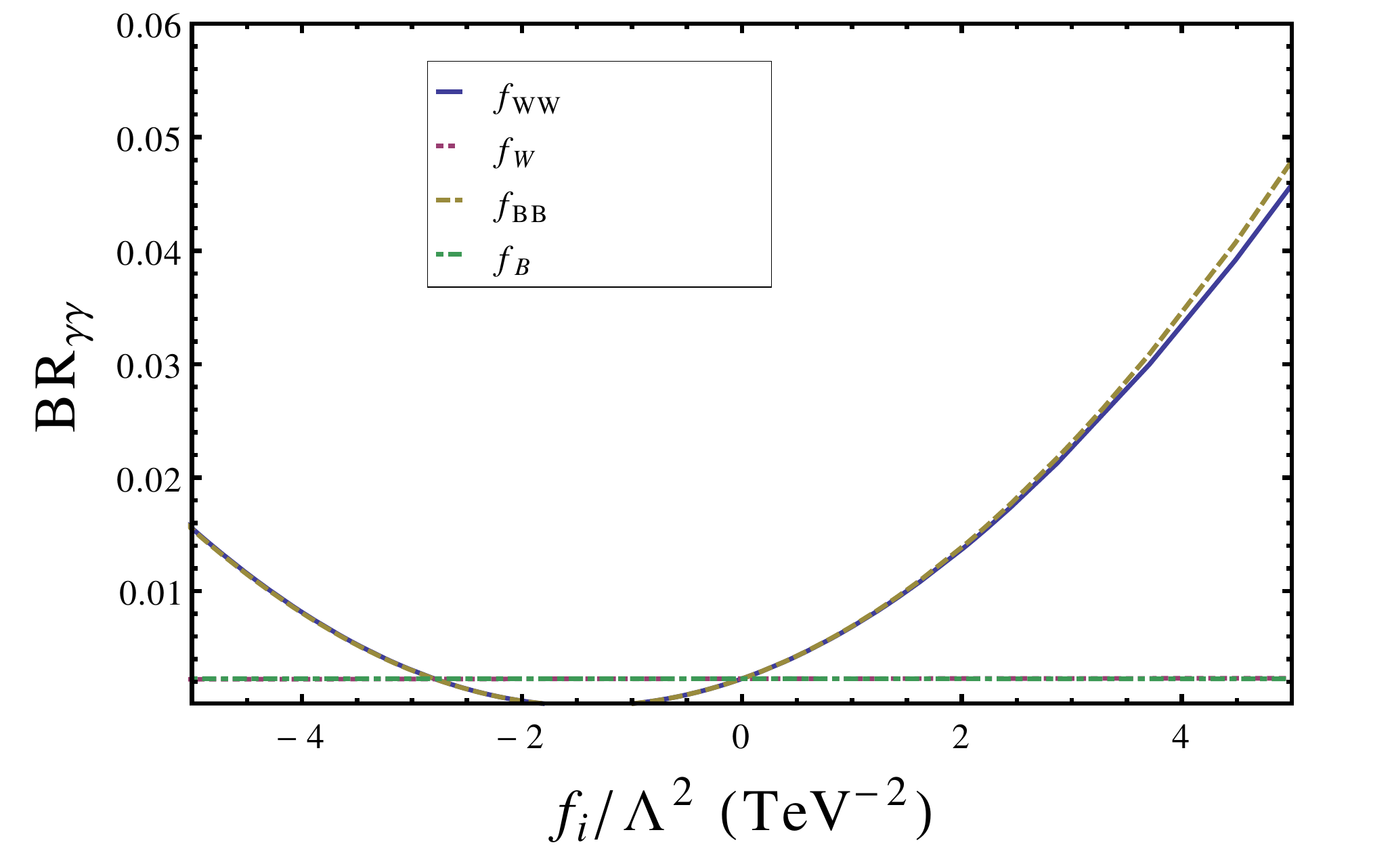}\label{fig:BRyy}}~~~
\subfloat[]{\includegraphics[width=6cm,height=6cm]{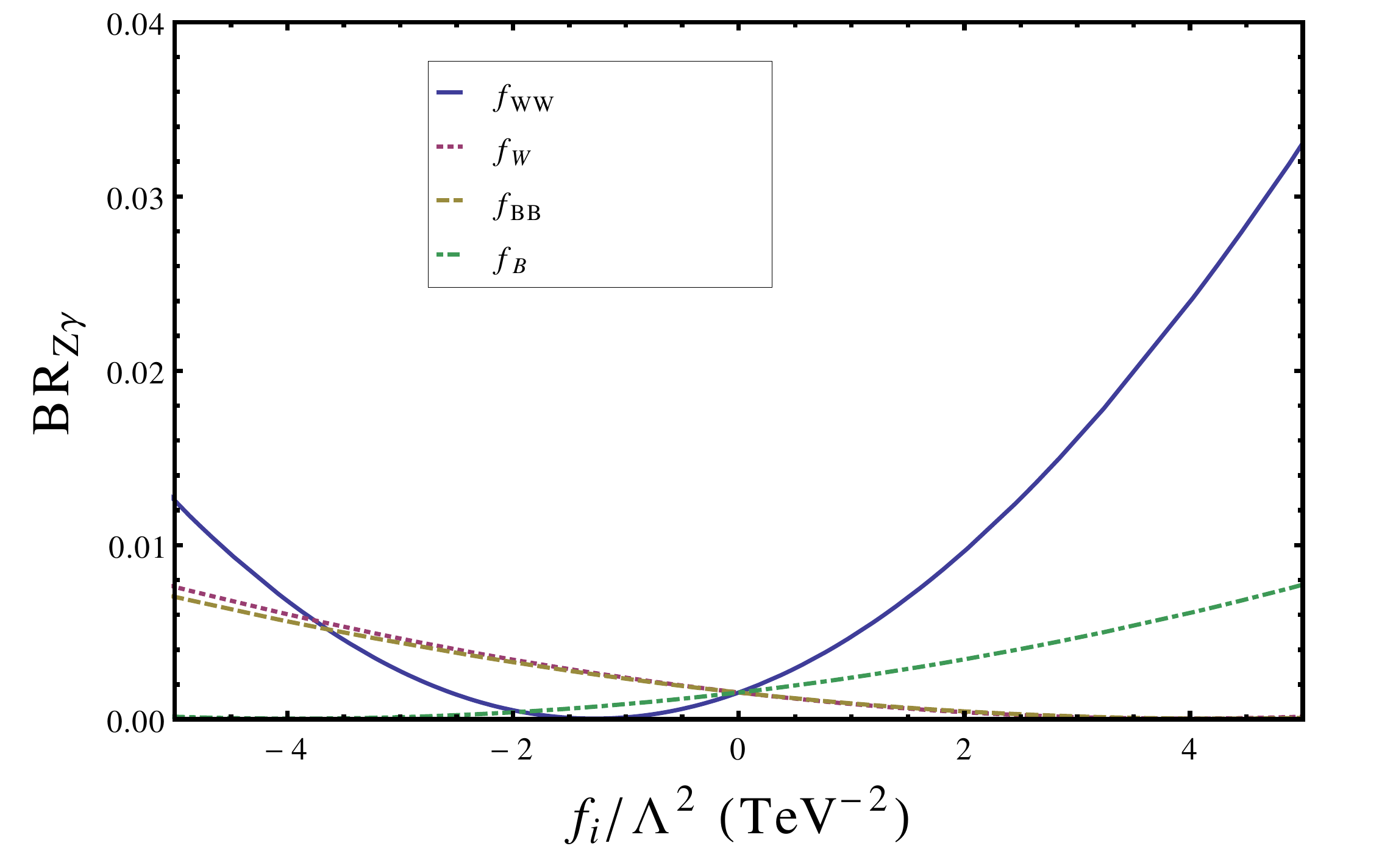}\label{fig:BRzy}}
\caption{Branching ratios of $H\to WW^*,ZZ^*,\gamma\gamma,Z\gamma$ in
  presence of HDOs. The operators are varied one at a time.}
\label{fig:BRHD}
\end{figure*}

The four HDOs under consideration affect Higgs
production as well as its decays, albeit to various degrees. For example, HDO-dependent single Higgs
production processes are in association with vector bosons ($VH$)
\textit{i.e.} $pp\to VH$ (where $V=\{W,Z\}$) and vector-boson fusion
($VBF$).  We show the production cross-sections in these channels at 14
TeV in Fig.~\ref{fig:Hprod}, as functions of the four operator
coefficients ($f_i$) taken one at a time.\footnote{We have used
  CTEQ6L1 parton distribution functions (PDFs) by setting the
  factorization  ($\mu_F$) and renormalization scales ($\mu_R$) at
  the Higgs mass ($M_H=125$ GeV).}
The relevant decay channels which are dependent on such
operators are $H\to WW^*,ZZ^*,\gamma\gamma,Z\gamma$.  
 Fig.~\ref{fig:BRHD} contains these branching ratios (BR) as functions
of the four coefficients under consideration.

The $VBF$ and $VH$
rates are  sensitive to  $f_{WW}$ and $f_W$, but depend very weakly
on $f_{BB}$ and $f_B$, while the cross-section $\sigma(pp\to WH)$,
 is completely independent of $f_{BB}$ and $f_B$).  The
HDO effects in $H\to \gamma\gamma$ and $H\to Z \gamma$ for $f_i\sim \mathcal{O}(1)$
\footnote{If the operators arise from loop-induced diagrams which imply
`loop factors' in denominators of the effective interactions, 
    $O(1)$ TeV$^{-2}$ coefficients imply strongly coupled
    theories~\cite{Elias-Miro:2013mua,Einhorn:2013kja}. However, if
    such operators originate from tree-level diagrams, then $O(1)$
    TeV$^{-2}$ coefficients imply weakly-coupled theories.}  is of
the same order as the loop-induced SM contribution unlike in the case
of the $HWW$ and $HZZ$ couplings. Therefore, $\textrm{BR}_{H\to
  \gamma\gamma}$ becomes highly sensitive to $f_{WW}$ and
$f_{BB}$. Consequently, the 7$+$8 TeV data already restrict their
  magnitudes. Bounds on all these operators in a similar framework can
  be seen in Table VI of Ref.~\cite{Corbett:2012ja} and also in
  Ref.~\cite{Masso:2012eq}. In Ref.~\cite{Corbett:2012ja}, the bounds
  have been presented at 90\% CL by varying multiple operators at the
  same time. These bounds have been obtained by considering the LHC
  data as well as constraints from on the oblique parameters,
  \textit{viz.}, $S,T$ and $U$. Bounds coming from the oblique
  parameters are generally weaker than those obtained from the LHC
  data as can be seen in Ref.~\cite{Masso:2012eq}. These limits may
  not be applicable when the analysis is performed varying one
  operator at a time.

Based on the above information, we set out to find observables
which are sensitive to $f_i \lesssim 5$ TeV$^{-2}$ in the High luminosity run at the LHC.
It is not completely clear yet how much of statistics is required to probe 
such small values with various event shape variables. On the other hand, 
the more straightforward observables, namely, total rates in various channels,
are also fraught with statistical, systematic and theoretical uncertainties which
must be reduced as far as possible when precision is at a premium.

An approach that is helpful is looking at ratios of cross-sections in
different channels. In this paper, we invoke
two kinds of ratios. First, we take ratios of events in two different
final states arising from a Higgs produced via the same channel
(in our case, gluon fusion). Such a ratio enables one to get rid
of correlated theoretical uncertainties (CThU) such as those in
PDF and renormalisation/factorisation
scales. They also cancel the uncertainty in total width which is
 correlated in the calculation of BRs into the two final states.
Secondly,  we consider the ratio of rates for the same final state
for two different production channels (such as $VBF$ and $VH$). 
Although the uncertainty in the BR cancels here,
the theoretical uncertainties at the production level do not. Moreover,
since the final state is same in this case, some systematic uncertainties which are
correlated (related to identification, isolation, trigger etc.) will also get cancelled.
However, this is helpful in another manner. For some of
the operators, the $f_i$-dependent shifts with respect to the SM
are in opposite direction for the numerator and the denominator in such
ratios. The result is that the net deviation adds up, as shown in 
subsection~\ref{subsec:R2}. We shall see that the use of both these kinds of ratios
(including those involving the channel $Z\gamma$ can capture
the HDO coefficients at a level unprecedented, going down to
values where new physics can show up.

\subsection{Observable sensitive to $\mathcal{O}_{WW}$ and $\mathcal{O}_{BB}$: $\mathcal{R}_1$}

As has been noted earlier, $\tr{BR}_{H\to\gamma\gamma}$ (Fig.~\ref{fig:BRyy})
is highly sensitive to two of the operators, namely,
$\mathcal{O}_{BB}$ and $\mathcal{O}_{WW}$. Therefore, we propose to
probe them in the $\gamma\gamma$ channel, with the Higgs produced
through gluon-gluon fusion ($ggF$). This final state is clean for reconstruction,
and has high statistics. We should mention here that if we consider
the simultaneous presence of more than one operators, then there is a
``blind-direction'' in the parameter space $f_{WW}\approx -f_{BB}$
where $\tr{BR}_{H\to \gamma\gamma}$ mimics the SM value.  This is because
the higher-dimensional part of the $H\gamma\gamma$ vertex is
proportional to $f_{WW}+f_{BB}$. Also, for the non-trivial range
$f_{WW}=f_{BB}\approx -3$, $\tr{BR}_{H\to \gamma\gamma}$ mimics the SM
value, due to parabolic dependence of the diphoton rate on the HDO
coefficients.  Therefore, the Higgs produced through $ggF$ followed by
its decay to $\gamma\gamma$ cannot be used alone to probe these two 
`special' regions of the parameter space.
We construct the  observable
\begin{equation}
\mathcal{R}_1(f_i)=\frac{\sigma_{\tr{ggF}}\times \tr{BR}_{H\to \gamma
    \gamma}(f_i)}{\sigma_{\tr{ggF}}\times \tr{BR}_{H\to WW^* \to 2 \ell 2 \nu}(f_i)},
\end{equation}
where $\ell=e,\mu$ and $f_i$'s are the operator coefficients.
As explained earlier, the CThU in production as well as total
width cancels here; so does the $K$-factor in the production rate.
Clearly, $\mathcal{R}_1$ can also be expressed as the ratio of
two signal strengths as follows,
\begin{equation}
\mathcal{R}_1(f_i) = \frac{\mu^{\tr{ggF}}_{\gamma\gamma}(f_i)}{\mu^{\tr{ggF}}_{WW^*}(f_i)} \times
\frac{(\sigma_{\tr{ggF}}\times \tr{BR}_{H\to \gamma
\gamma})^{\tr{SM}}}{(\sigma_{\tr{ggF}}\times \tr{BR}_{H\to WW^* \to 2 \ell 2 \nu})^{\tr{SM}}}\ .
\end{equation}

Therefore, already measured $\gamma\gamma$ and $WW^*$ signal strengths can be used to constrain the operator coefficients affecting the
ratio $\mathcal{R}_1$. 
The efficiency of acceptance cuts does not affect the results,
for values of $f_{WW}$ and $f_{BB}$ which are of relevance here because for such small values 
of the parameter coefficients the change in experimental cut-efficiencies is negligible.
On top of that, for the $ggF$ production mode, these operators only affect the decay 
vertices and hence the cut-efficiencies are but modified by a very small extent. We must 
also note that in defining $\mathcal{R}_1$ a full jet-veto (0-jet category) has been demanded for both 
the numerator and the denominator to reduce the uncertainties related to the
different jet-requirement in the final state. Besides, in the denominator, the $WW^*$ pair is considered 
to decay into both same flavour ($ee+\mu\mu$) and different flavour ($e\mu + \mu e$) final states to improve the statistics.

\subsection{Observable sensitive to $\mathcal{O}_{WW}$ and $\mathcal{O}_{W}$: 
$\mathcal{R}_2$}
\label{subsec:R2}

It turns out that the $f_{WW}$ and $f_W$ affect (to one's advantage)
the ratio of events in a particular Higgs decay mode in the $VBF$ and $VH$
channels.  This captures the new physics at the production level.  By
considering the same final states from Higgs decay, some theoretical
uncertainties in the decay part cancels out. The production level
uncertainties, including the $K$-factors, however, do not cancel
here. In our calculation, the next-to-next-to leading order (NNLO) $K$-factors
have been assumed to be
the same as in the SM, expecting that the presence of HDO does not effect the $K$-factors much. For precise estimate of the observed ratio, one
of course has to incorporate the modified cut efficiencies due to the
new operators, though such modifications may be small. The other,
important advantage in taking the above kind of ratio is that, for
not-too-large $f_{WW}$ or $f_W$ (in the range $[-5,+5]$), the deviations of the $VBF$ and
$VH$ cross-sections are in opposite directions. The generic
deviation for the rate in any channel can be parametrized as
\begin{equation} 
\sigma^{\textrm{HDO}}_{\textrm{prod.}}=\sigma^{\textrm{SM}}_{\textrm{prod.}}\times \left(1+\delta_{\textrm{prod.}}\right).
\end{equation}
From Fig.~\ref{fig:VBF}, $\delta_{\textrm{VBF}}$ is positive in the
range $f_{WW},f_W>0$. On the other hand, in the same region of the
parameter space, $\delta_{\textrm{VH}}$ is negative as evident from
Figs.~\ref{fig:WH} and~\ref{fig:ZH}.  Hence, on taking the ratio
$\sigma_{\textrm{VBF}}^{\textrm{HDO}}/\sigma_{\textrm{VH}}^{\textrm{HDO}}$, the deviation from
SM is
\begin{equation}
\frac{\sigma_{\textrm{VBF}}}{\sigma_{\textrm{VH}}}
=\frac{\sigma_{\textrm{VBF}}^{\textrm{SM}}}{\sigma_{\textrm{VH}}^{\textrm{SM}}}\times
\left(1 + \delta_{\textrm{VBF}} - \delta_{\textrm{VH}} +
\mathcal{O}({\delta^{2}})\right).
\end{equation}
Thus this ratio further accentuates the deviation from SM behaviour.
As an example, if we consider the parameter choice $f_W=2$, then
$\delta_{\tr{VBF}}\approx 3.6$\% and $\delta_{\tr{WH}}\approx 10$\%. However,
from the ratio, the combined $\delta_{\tr{VBF+WH}} \approx 15$\%, which is
a clear indication of why we should consider such ratios. We thus
define our next observable
\begin{equation}
\label{eq:R2}
\mathcal{R}_2(f_i)=\frac{\sigma_{\tr{VBF}}(f_i)\times \tr{BR}_{H\to \gamma
    \gamma}(f_i)}{\sigma_{\tr{WH}}(f_i) \times \tr{BR}_{H\to \gamma \gamma}(f_i)
  \times \tr{BR}_{W \to \ell \nu}},
\end{equation}
where the $\gamma \gamma$ final state has been chosen because of its
clean character and reconstructibility of the Higgs mass.  It should be remembered, 
however, that $f_{WW}, f_{BB}$ in the range $-3$ to 0  causes the diphoton
branching ratio to undergo a further dip. This can adversely affect
the statistics, and thus the high luminosity run is required for an exhaustive
scan of the admissible ranges of the above coefficients.

\subsection{Observable sensitive to $\mathcal{O}_{B}$: $\mathcal{R}_3$}
\label{subsec:R3}
The operator $\mathcal{O}_{B}$ is sensitive to $H\to ZZ^*$ and $H\to
Z\gamma$. In the former mode, the sensitivity of $f_B$ is limited (see
the green curve in Fig.~\ref{fig:BRzz}) and can be appreciable only
for larger $f_B$. The partial decay width $\Gamma_{H\to Z\gamma}$, on
the other hand is rather sensitive to all the four operators under
study (Fig.~\ref{fig:BRzy}), primarily due to the fact that the new
$HZ\gamma$ vertex contributes practically as the same order as in the
SM. However, the present statistics in this channel is
poor~\cite{Aad:2014fia,Chatrchyan:2013vaa}. We expect better
  bounds on $\mathcal{O}_{WW}$, $\mathcal{O}_{BB}$ and $\mathcal{O}_W$
  from the measurements of $\mathcal{R}_1$ and $\mathcal{R}_2$.  We
use $\mathcal{R}_3$ for the 14 TeV 3000 fb$^{-1}$ run to constrain
$f_B$ only, for which other channels fail.  In the same spirit as for
$\mathcal{R}_1$, we thus define our third observable
\begin{equation}
\mathcal{R}_3(f_i)=\frac{\sigma_{\tr{ggF}}\times \tr{BR}_{H\to Z
    \gamma \to 2 \ell \gamma}(f_i)}{\sigma_{\tr{ggF}}\times \tr{BR}_{H\to WW^* \to 2 \ell 2 \nu}(f_i)},
\end{equation}
where $\ell=e,\mu$ and here again the CThU cancels. Here also, we must 
note that in defining $\mathcal{R}_3$ a full jet-veto has been demanded for both 
the numerator and the denominator. For the numerator, the $Z$ boson's  
decay to both an electron pair and a muon pair is considered. Besides, in the denominator, the 
$WW^*$ pair is taken to decay similar to the $\mathcal{R}_1$ case.

\textbf{Comparison with the $\kappa$-framework :} In principle,
  studies in terms of ratios in different channels can be carried also
  in the $\kappa$-framework~\cite{Banerjee:2012xc,Heinemeyer:2013tqa,Khachatryan:2014jba,ATLAS-CONF-2015-007} in which couplings are
  modified just by scale factors. It should, however, be remembered
  that the present analysis involves new Lorentz structures and hence
  brings non-trivial interference terms in the squared
  amplitudes. Unlike the situation with overall scaling, this prevents
  the cancellation of the modifying couplings when one considers
  ratios of events taking (SM+BSM) effects into account.

  Even though the ratio $\mathcal{R}_1$ ($\mathcal{R}_3$), dominated
  by $H \gamma\gamma$ ($H Z \gamma$) vertex, contains no new Lorentz
  structures, it is still sensitive to the HDOs due to the presence of
  the $HWW$ vertex in the denominator. Therefore, these ratios,
  although apparently similar to ratios employing the
  $\kappa$-framework, are different in practice.  $\mathcal{R}_2$ is a
  ratio of $\sigma_{VBF}$ and $\sigma_{WH}$ which are sensitive to the
  operator coefficients as shown in Fig.~\ref{fig:Hprod}. In the
  $\kappa$-framework, $\sigma_{VBF}$ is dominated by the $WWH$ vertex
  and hence $\kappa_{WW}$ will approximately cancel in
  $\mathcal{R}_2$. On the other hand, there will be no trivial
  cancellations between the numerator and denominator in the
  HDO-framework.

\section{Results of the analysis}
\label{results}

For our subsequent collider analysis, the chain we have used is as follows - 
first we have implemented the relevant dimension-6 
interaction terms as shown in Eq.~(\ref{eq:lagHVV}) in~\textsc{FeynRules}~\cite{Alloul:2013bka}, 
and generated the Universal FeynRules Output (UFO)~\cite{Degrande:2011ua} model files. These UFO 
model files have been used in the~\textsc{Monte-Carlo} (MC) event generator~\textsc{MadGraph}~\cite{Alwall:2014hca} to generate event samples. Next, the parton-showering and hadronisation are performed using 
\textsc{Pythia}~\cite{Sjostrand:2006za} and
finally detector level analyses is carried using
\textsc{Delphes}~\cite{deFavereau:2013fsa}.

Before we discuss the phenomenological aspects of the aforementioned
observables, we re-iterate below the various kinds of
uncertainties considered. The two major classes of
observables where these uncertainties arise are as follows:

\begin{itemize}
\item \underline{Same production channel but different final states:}\\ 
  In such cases  (as in $\mathcal{R}_1$ and $\mathcal{R}_3$), the
  correlated uncertainties lie in  PDF+$\alpha_s$, QCD-scale and in
  the total Higgs decay width, $\Gamma_H$. However,
  uncertainties in the partial decay widths are uncorrelated~\footnote{We must mention here that $\Gamma_{H \to \gamma \gamma}$ and 
  $\Gamma_{H \to Z \gamma}$ have tiny correlations with $\Gamma_{H \to WW^*}$ because of the $W$-boson loop in the former two cases. 
  However, in this present analysis we neglect such small correlations and consider these partial decay widths to be mostly 
  uncorrelated}. 
  Statistical uncertainties for distinct final states are always uncorrelated
  and are retained in our analysis.  
  We also assume some systematic uncertainties, whenever shown, to be  
  fully uncorrelated. All surviving uncertainties are added in quadrature
  to estimate total uncertainties related to our observables.

\item \underline{Different production channels but same final state:}\\ For
  such observables ($\mathcal{R}_2$ in our definition), the only correlated
  uncertainty is in $\tr{BR}_{H\to \gamma \gamma}$. All other uncertainties
  are uncorrelated and hence are added in quadrature (including the
  uncertainties in the numerator and the denominator of the ratio $\mathcal{R}_2$). Beside the
  already mentioned theoretical uncertainties, we also encounter some
  additional theoretical uncertainty related to the QCD-scale in the $WH$ mode, which we separately discuss 
  in  subsection~\ref{subsec:R214}.
\end{itemize}

\begin{table}
\begin{center}
\begin{tabular}{|c|c|c|c|}
\hline 
SM Quantity & Value & +\textit{ve} uncert. \% & $-$\textit{ve} uncert. \% \\ 
\hline 
$\tr{BR}_{H\to\gamma\gamma}$ & $2.28\times 10^{-3}$ & $+4.99$ & $-4.89$ \\ 
\hline 
$\tr{BR}_{H\to WW^*}$ & $2.15\times 10^{-1}$  & $+4.26$ & $-4.20$ \\ 
\hline 
$\tr{BR}_{W\to e \nu_e}$ & $1.07\times 10^{-1}$ & $+0.16$ & $-0.16$ \\
\hline
$\tr{BR}_{W\to \mu \nu_{\mu}}$ & $1.06\times 10^{-1}$ & $+0.15$ & $-0.15$ \\
\hline
$\tr{BR}_{H\to Z \gamma}$ & $1.54\times 10^{-3}$ & $+9.01$ & $-8.83$ \\
\hline
$\tr{BR}_{Z\to ee}$ & $3.36 \times 10^{-2}$ & $+0.004$ & $-0.004$ \\
\hline
$\tr{BR}_{Z\to \mu \mu}$ & $3.37 \times 10^{-2}$ & $+0.007$ & $-0.007$ \\
\hline
Total $\Gamma_H$ & $4.07$ MeV & $+3.97$ & $-3.94$  \\ 
\hline
\end{tabular}
\caption{$\tr{BR}_{H\to\gamma\gamma}$, $\tr{BR}_{H\to WW^*}$, $\tr{BR}_{H\to Z \gamma}$, $\tr{BR}_{W\to \ell \nu}$, $\tr{BR}_{Z\to \ell \ell}$ and total Higgs width 
  $\Gamma_H$ (MeV) and their \% uncertainties ($+ve$ and $-ve$ refer to positive and negative uncertainties respectively)} for a Higgs 
  of mass 125 GeV ($m_W = 80.385$ GeV and $m_Z = 91.1876$ GeV). These numbers are taken from the LHC Higgs Cross Section Working Group 
  page~\cite{twiki}.
\label{tab:BR-width-SM}
\end{center}
\end{table}
\begin{table}
\centering
\begin{tabular}{|c|c|c|c|c|c|}
\hline Process & $\sigma$ (pb) & \footnotesize{$+$QCD-Scale \%} & \footnotesize{$-$QCD-Scale \%} &
\footnotesize{$+$(PDF$+\alpha_S$) \%} & \footnotesize{$-$(PDF$+\alpha_S$) \%} \\ \hline $ggF$ & 49.47 &
$+7.5$ & $-8.0$ & $+7.2$ & $-6.0$ \\ 
\hline 
$VBF$ & 4.233 & $+0.4$ & $-0.5$ & $+3.3$ & $-3.3$ \\ 
\hline 
$WH$ & 1.522 & $+0.8$ & $-1.6$ & $+3.2$ & $-3.2$ \\ 
\hline 
$ZH$ & 0.969 & $+4.0$ & $-3.9$ & $+3.5$ & $-3.5$ \\ 
\hline
\end{tabular}
\caption{The cross-sections of relevant Higgs production ($m_H=125$ GeV) channels
and their QCD-Scale and PDF$+\alpha_s$ uncertainties in \%. These numbers are again taken from the LHC Higgs Cross Section Working Group page~\cite{twiki}.}
\label{tab:sigma-SM}
\end{table}
\begin{table}
\centering
\begin{tabular}{|c|c|c|c|}
\hline 
 & $\mathcal{R}_1$ & $\mathcal{R}_2$ & $\mathcal{R}_3$ \\ 
\hline 
$N_S^{num}$ & 47724 ($\gamma\gamma$ in $ggF$) & 194 ($\gamma\gamma$ in $VBF$) & 1989 ($Z\gamma$ in $ggF$) \\ 
\hline 
$N_B^{num}$ & $3.16\times 10^6$ & 1041 & 691931 \\
\hline
$N_S^{den}$ & 40850 ($WW^*$ in $ggF$) & 238 ($\gamma\gamma$ in $WH$) & 40850 ($WW^*$ in $ggF$) \\ 
\hline
$N_B^{den}$ & 366450 & 995 & 366450 \\
\hline
\end{tabular} 
\caption{Number of surviving events (taken from Refs.~\cite{ATLAS-HL-LHC,ATLAS-HL-LHC-gaga}) after the selection cuts in the SM at 14 TeV with 
3000 fb$^{-1}$ integrated luminosity. These numbers are used to compute the statistical
uncertainties (which goes as $\sqrt{N_S+N_B}/N_S$, where $N_S$ and $N_B$ are respectively the number of surviving signal and background 
events after selection cuts) related to the numerator and denominator of the three observables. Number of events in the $VBF$ ($\gamma\gamma$) 
channel is computed by applying a fixed $p_T$-cut (keeping other cuts same as in Ref.~\cite{ATLAS-HL-LHC}) of 50 GeV on both the 
tagged jets instead of $\eta$-dependent jet selection cuts as used in the same reference. Number of events for 
$\gamma \gamma$ in $\mathcal{R}_1$, $Z \gamma$ in $\mathcal{R}_3$ and $WW^*$ for $\mathcal{R}_1$ and $\mathcal{R}_3$ are obtained after 
putting 0-jet veto and demanding only $ggF$ events. The superscripts $num$ and $den$ signifies the numerators and denominators of the 
three observables.}
\label{tab:Nev}
\end{table}
\begin{table}
\centering
\begin{tabular}{|c|c|c|}
\hline 
$\mathcal{R}_1$ & $\mathcal{R}_2$ & $\mathcal{R}_3$ \\ 
\hline 
2.87 \% & 13.83 \% & 29.63 \% \\ 
\hline
\end{tabular} 
\caption{Statistical uncertainty for the observables $\mathcal{R}_1$, $\mathcal{R}_2$ and $\mathcal{R}_3$. 
The numbers are obtained after doubling the number of signal and background events given in Table~\ref{tab:Nev} in order to
account for both ATLAS and CMS experiments.}
\label{tab:stat}
\end{table}
\begin{table}
\centering
\begin{tabular}{|c|c|c|c|}
\hline 
 & $\mathcal{R}_1$ & $\mathcal{R}_2$ & $\mathcal{R}_3$ \\ 
\hline 
Numerator & 2.5\% ($\gamma\gamma$ in $ggF$) & 9.1\% ($\gamma\gamma$ in $VBF$) & 3.1\% ($Z\gamma$ in $ggF$) \\ 
\hline 
Denominator & 3.4\% ($WW^*$ in $ggF$) & 5.0\% ($\gamma\gamma$ in $WH$) & 2.8\% ($WW^*$ in $ggF$) \\ 
\hline 
\end{tabular} 
\caption{Systematic uncertainties used in our analysis to compute the total uncertainties
related to the three observables. The numbers shown here are combination of various types of relevant systematic uncertainties added in quadrature taken from Refs.~\cite{Aad:2014eha,ATLAS:2014aga,Aad:2014fia}.}
\label{tab:systU}
\end{table}
We further assume that the percentage uncertainties remain same even
after the inclusion of the anomalous couplings. In order to
illustrate, how the uncertainties are taken into consideration, we
list the theoretical uncertainties related to relevant Higgs BR and
total width in Table~\ref{tab:BR-width-SM}, and related to various
production cross-sections in Table~\ref{tab:sigma-SM}. In
Table~\ref{tab:Nev}, we present the number of surviving events after
the selection cuts in the SM at 14 TeV with 3000 fb$^{-1}$ integrated
luminosity in the pure production modes. These numbers are taken
  from Refs.~\cite{ATLAS-HL-LHC,ATLAS-HL-LHC-gaga} except for the
  $\gamma\gamma$ channel in the $VBF$ production mode, which we
  estimate by applying a fixed $p_T$-cut (keeping other cuts are same
  as in Ref.~\cite{ATLAS-HL-LHC}) of 50 GeV on both the tagged jets
  instead of $\eta$-dependent jet selection cuts as used in the same
  reference. The number of events have been computed by removing
  the contaminations from other production mechanisms which will reduce the number of events and hence
  enhance the statistical uncertainties (which roughly goes as $\sqrt{N_S+N_B}/N_S$,
  with $N_S$ and $N_B$ being respectively the number of surviving signal and background 
  events after selection cuts). For instance, the reported number of $\gamma\gamma$ events
  for an integrated luminosity of 3000 fb$^{-1}$ is $49200$ with a
  $3\%$ contamination from $VBF$ (Table 3 in
  Ref.~\cite{ATLAS-HL-LHC}). In our analysis we have used $N_S=47724$
  ($=0.97 \times 49200$) to compute the statistical
  uncertainty. Similarly a $30\%$ contamination in the $VBF$ category
  due to $ggF$ (Table 3 in Ref.~\cite{ATLAS-HL-LHC}) has also been
  taken into consideration. In doing so, we are giving conservative estimates on
  the statistical uncertainties. All entries in Table~\ref{tab:Nev}
  are shown after removing contamination to compute conservative
  statistical uncertainties.  We must note that, while computing the
  statistical uncertainties (as shown in Table~\ref{tab:stat}) for all the three ratios, we double the
  number of events in Table~\ref{tab:Nev} to roughly accommodate two
  independent experiments to be performed by ATLAS and CMS. Here, we
  assume that ATLAS and CMS will analyse the same channels with
  similar set of selection cuts and will roughly obtain same number of
  events in the actual experiment. It is also assumed that the overall 
  performance of ATLAS and CMS will be similar, integrated over a large luminosity. 
  In future, when the data become actually available, one would be able to compute 
  the exact statistical uncertainties. However, we must note that one should actually
  take the number of events in the \textit{side-band} ($N_{side-band}$) in order to compute the statistical
  uncertainties. The procedure we follow gives conservative values for the statistical
  uncertainties. In future, the actual experiments will provide us $N_{side-band}$ which
  will allow us to compute accurate statistical uncertainties. However, the \textit{side-band} analysis is beyond the scope of this paper as the data for the 14 TeV run at 3000 fb$^{-1}$ is yet unavailable.

We also use some systematic uncertainties in our analysis as listed in Table~\ref{tab:systU} (Refs.~\cite{Aad:2014eha,ATLAS:2014aga,Aad:2014fia}). 
In the future, it is quite expected, various systematic uncertainties will reduce
by improving their modelling. To be conservative, we have used various important uncorrelated
systematic uncertainties as used in Refs.~\cite{Aad:2014eha,ATLAS:2014aga,Aad:2014fia} for 7+8 TeV analysis. For the 
observable $\mathcal{R}_1$, since we are applying same jet veto (\textit{i.e.} 0-jet
category), the systematic uncertainties related to the jet energy scale, jet 
vertex fraction etc. will not be present. On the other hand, due to the different
final state, systematic uncertainties related to the photon and lepton identification and isolation, missing energy trigger etc. will remain. In a similar
fashion, for $\mathcal{R}_2$ and $\mathcal{R}_3$ various correlated systematic uncertainties will cancel between their respective numerator and denominator.

Next, we consider the
ratio $\mathcal{R}_1$ in the light of both the existing data and those predicted
for the high energy run. For $\mathcal{R}_2$ and $\mathcal{R}_3$, only a discussion in
terms of 14 TeV rates is relevant, as the currently available results
have insufficient statistics on these.

\subsection{$\mathcal{R}_1$ @ 7$+$8 TeV}

Before predicting the bounds from the 14 TeV HL run, let us form an
idea about the constraints from the 7+8 TeV Higgs data in the
$\gamma\gamma$ and $WW^*$ channels. In Table~\ref{tab:mu-gagaWW}, we show the \emph{exclusive} signal strengths in the $\gamma\gamma$ and $WW^*$ final states through the
  $ggF$ production mode as reported by
  ATLAS~\cite{Aad:2014eha,ATLAS:2014aga} and
  CMS~\cite{Khachatryan:2014ira,CMS:2014ega}.

We must emphasize that the categorization introduced by the ATLAS and CMS experiments are used to enhance the sensitivity for the Higgs boson signal (Tables II and III in Ref.~\cite{Aad:2014eha}). The signal strengths ($\mu$) shown in Fig. 17 include these contaminations. These signal strengths are further combined to give specific production categories as shown in Fig. 18. For instance $\mu$ for \textit{ggF categories} is the combination of the four categories, \textit{viz.} central low $P_{T_t}$, central high $P_{T_t}$, forward low $P_{T_t}$ and forward high $P_{T_t}$. Therefore, the $\mu$ for specific categories in Fig. 18 is not \emph{exclusive}. However, while obtaining the $\mu$ for a specific production mode in Fig. 19, the effect of contaminations are properly removed (by knowing the amount of contaminations from Monte-Carlo simulation for the SM) and therefore, these are the \emph{exclusive} signal strengths. The removal of contaminations includes not only the subtraction of production mechanisms that are not of interest but also the propagation of errors. The experiments have taken into account the impact on the statistical, systematic and theoretical errors for the extraction of the \emph{exclusive} signal strengths. Therefore, the \emph{exclusive} $\mu$ will generally contain larger uncertainty. For example one can see that the error on the global signal strength is significantly better than that extracted for individual production mechanisms. For instance, in Ref.~\cite{Aad:2014eha}, where ATLAS reports on signal strengths with the di-photon channel, the global signal strength is $\mu = 1.17 \pm 0.27$, which leads to an accuracy of 23\%, whereas for the signal strength of gluon-gluon fusion (ggf) $\mu_{ggf} = 1.32 \pm 0.38$, corresponding to an accuracy of 29\%. Same applies to the results reported by CMS in Ref.~\cite{Khachatryan:2014ira}.

   Here we statistically combine the signal strengths for a
  particular final state as reported by the two experiments, using the
  following relations
\begin{equation}
\frac{1}{\bar{\sigma}^2}=\sum_{i} \frac{1}{\sigma_i^2};~~~~~ 
\frac{\bar{\mu}}{\bar{\sigma}^2}=\sum_{i} \frac{\mu_i}{\sigma_i^2},
\label{eq:comb}
\end{equation}
where $\bar{\sigma}$ ($\bar{\mu}$) refers to the combined $1\sigma$
uncertainty (signal strength) and $\sigma_i$ ($\mu_i$) signifies the
corresponding uncertainties (signal strengths) in different
experiments.

\begin{figure*}[ht!]
\centering
\subfloat[]{\includegraphics[width=7cm,height=7cm]{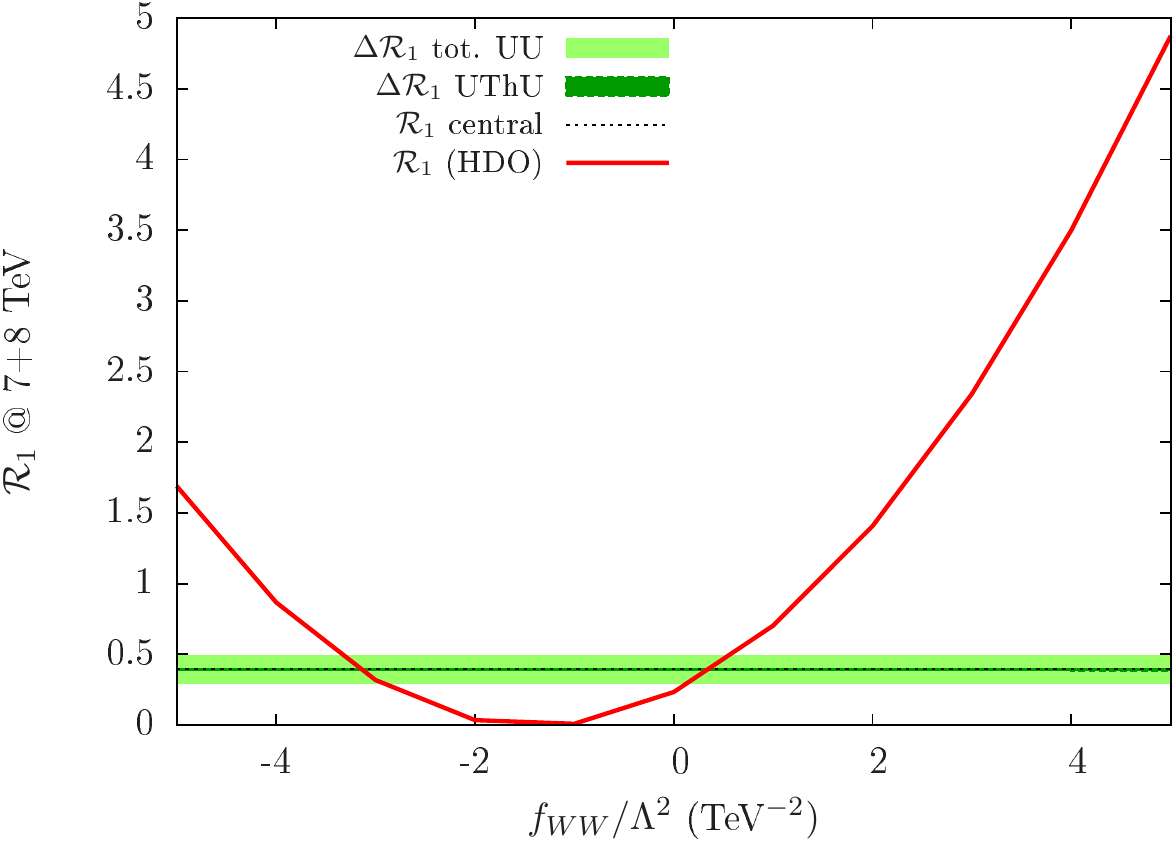}\label{fig:BRyywwfWW}}~~~
\subfloat[]{\includegraphics[width=7cm,height=7cm]{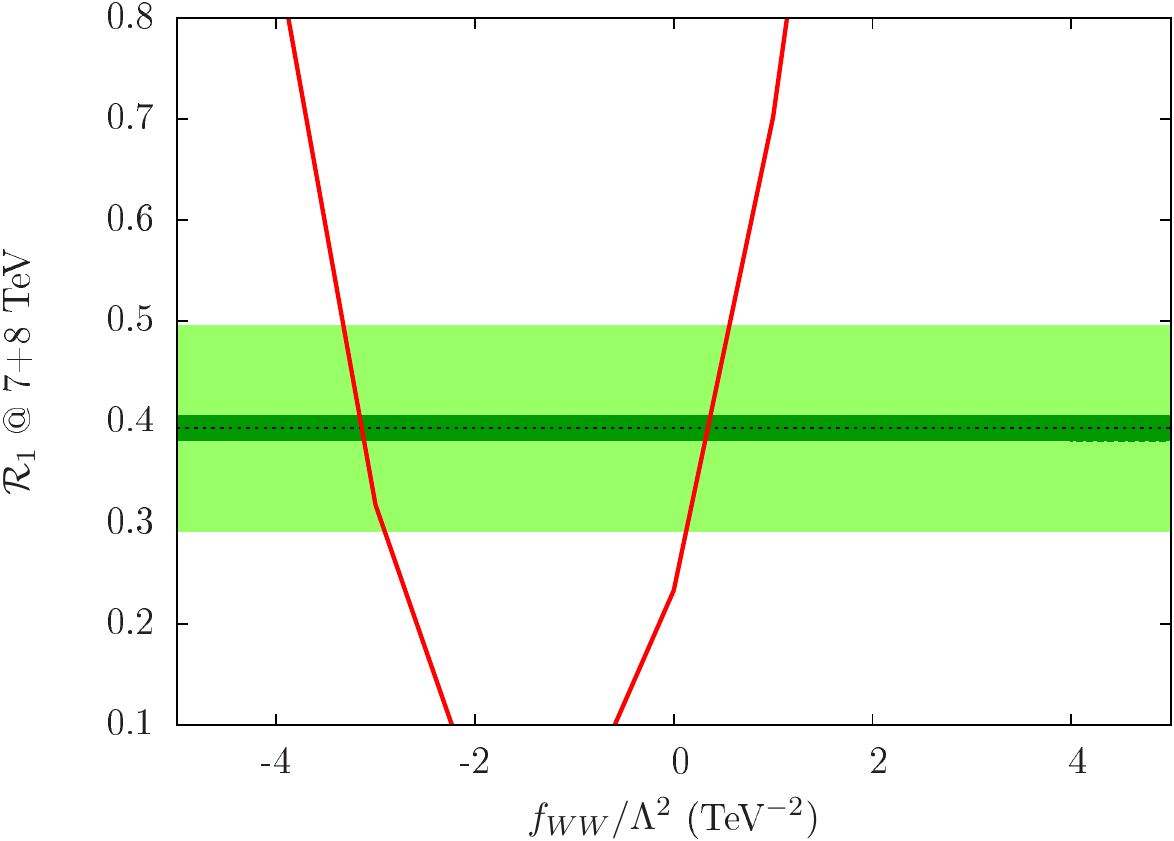}\label{fig:BRyywwfBB}}
\caption{(a) $\mathcal{R}_1$ versus $f_{WW}/\Lambda^2$ (TeV$^{-2}$) and (b) same
plot in magnified scale. Plots (a) and (b) are identical for $f_{BB}/\Lambda^2$. The red line is the theoretical
  expectation in presence of HDOs. The inner band (dark green) shows the uncorrelated theoretical 
  uncertainty (UThU) and the outer (light green) band shows the total
  surviving uncorrelated uncertainty (UU) (uncorrelated theoretical + statistical + systematic) at
  7+8 TeV computed using the $\mu_{\gamma \gamma}$ and $\mu_{WW^*}$
  (CMS+ATLAS) results. The black dotted line is the corresponding central
  value. The uncertainty bands correspond to 68\% CL.}
\label{fig:R18}
\end{figure*}

\begin{table}
\begin{center}
\begin{tabular}{|c|c|c|}
\hline Experiment & $\mu(H\to \gamma \gamma)$ in $ggF$ & $\mu(H \to
  WW^* \to 2\ell \slashed{E}_T)$ in $ggF$ \\ 
\hline 
ATLAS (@ 7+8 TeV) & $1.32^{+0.38}_{-0.38}$ & $1.02^{+0.29}_{-0.26}$ \\ 
\hline 
CMS (@ 7+8 TeV) & $1.12^{+0.37}_{-0.32}$ & $0.75^{+0.29}_{-0.23}$ \\ 
\hline
Combined & $1.21 \pm 0.26$ & $0.88 \pm 0.19$ \\
\hline
\end{tabular}
\caption{Measured Higgs Signal strengths in the $\gamma \gamma$ and $WW^*$
modes where Higgs is produced through only $ggF$ channel using $\sqrt{s} = 7 + 8$ TeV data by ATLAS~\cite{Aad:2014eha,ATLAS:2014aga} and CMS~\cite{Khachatryan:2014ira,Khachatryan:2014jba}. Here we have combined the ATLAS and
CMS signal strengths for a particular final state and production mode using Eq.~\ref{eq:comb}.}
\label{tab:mu-gagaWW}
\end{center}
\end{table}
We compute all the surviving correlated
theory errors and subtract them in quadrature from the errors in
the numerator and denominator of the ratio $\mathcal{R}_1$, \textit{viz.}
$\mathcal{R}_{1}^{num.}=\mu_{H\to \gamma \gamma}^{\tr{ggF}} \times
(\sigma_{\tr{ggF}}\times \tr{BR}_{H\to \gamma \gamma})^{\tr{SM}}$ and
$\mathcal{R}_{1}^{den.}=\mu_{H\to WW^*}^{\tr{ggF}} \times (\sigma_{\tr{ggF}}\times
\tr{BR}_{H\to WW^*})^{\tr{SM}} \times \sum_{\ell} \tr{BR}^2_{W\to \ell\nu_{\ell}}$~\footnote{For instance, the error associated with 
combined (ATLAS+CMS) $\mu^{ggF}(H\to \gamma \gamma)$ \textit{i.e.} $\pm 0.26$ consists of theoretical, statistical and systematic uncertainties and, by subtracting the CThU 
($\pm 0.13$) in quadrature we get ($\pm 0.22$) which will finally contribute to the 
uncertainty related to the numerator of $\mathcal{R}_1$.}. 
In Fig.~\ref{fig:R18}, the red line is the
theoretically computed $\mathcal{R}_1$ which is independent of the centre of
mass energy since $\mathcal{R}_1$ is actually a ratio of two BRs. 
The outer (light green) band shows the uncertainty comprising of the
uncorrelated theoretical, statistical and systematic parts and the inner (dark green)
band represents the total uncorrelated theory uncertainty. The black
dashed line gives the experimental central value of $\mathcal{R}_1$. 
The ratio, $\mathcal{R}_1$ is almost completely
dominated by $\tr{BR}_{H \to \gamma \gamma}$ (since $\tr{BR}_{H \to WW^*}$ is not so sensitive 
on HDOs) and therefore highly sensitive to the operators $\mathcal{O}_{WW}$ and
$\mathcal{O}_{BB}$. The parabolic nature of the $\tr{BR}_{H \to \gamma \gamma}$
as functions of $f_{WW}$ and $f_{BB}$ leads to two disjoint allowed ranges of 
$f_{WW}=f_{BB} \approx [-3.32,-2.91] \cup [0.12,0.57]$ as shown in 
Fig.~\ref{fig:R18}. We should mention that the region between these two allowed
ranges shows extremely low values of $\tr{BR}_{H\to \gamma\gamma}$ because of
destructive interference between the SM and HDO might leads to poor statistics.
If both $\mathcal{O}_{WW}$ and $\mathcal{O}_{BB}$ are present simultaneously
with almost equal magnitude and opposite signs, the observable $\mathcal{R}_1$
closely mimics the SM expectation, and to probe that `special' region of parameter space
we need to go for other observable like $\mathcal{R}_2$. The operators
$\mathcal{O}_{W}$ and $\mathcal{O}_{B}$ are mostly insensitive to this
observable mainly because $\tr{BR}_{\gamma\gamma}$ is independent of these
operators and the dependence of $\tr{BR}_{WW^*}$ on all four operators is
comparatively weak (see Fig.~\ref{fig:BRww})

We compare our results with the existing bounds on these operators as obtained in literature. For instance, the limits obtained in Fig. 3 (left panel) of Ref.~\cite{Masso:2012eq} on $\mathcal{O}_{WW} \, \textrm{and} \, \mathcal{O}_{BB}$ at 68\% CL are $[-3.23,-2.61] \cup [-0.35,0.27]$ (in TeV$^{-2}$) for the ATLAS case. In obtaining these limits, they varied one operator at a time. This is similar in approach to our study where we have given a framework where one operator is varied at a time. Our bounds are in very good agreement with their results. The slightly different limits obtained by us are due to the use of more recent data in our case.

\subsection{$\mathcal{R}_1$ @ 14 TeV}

\begin{figure*}[ht!]
\centering
\subfloat[]{\includegraphics[width=7cm,height=7cm]{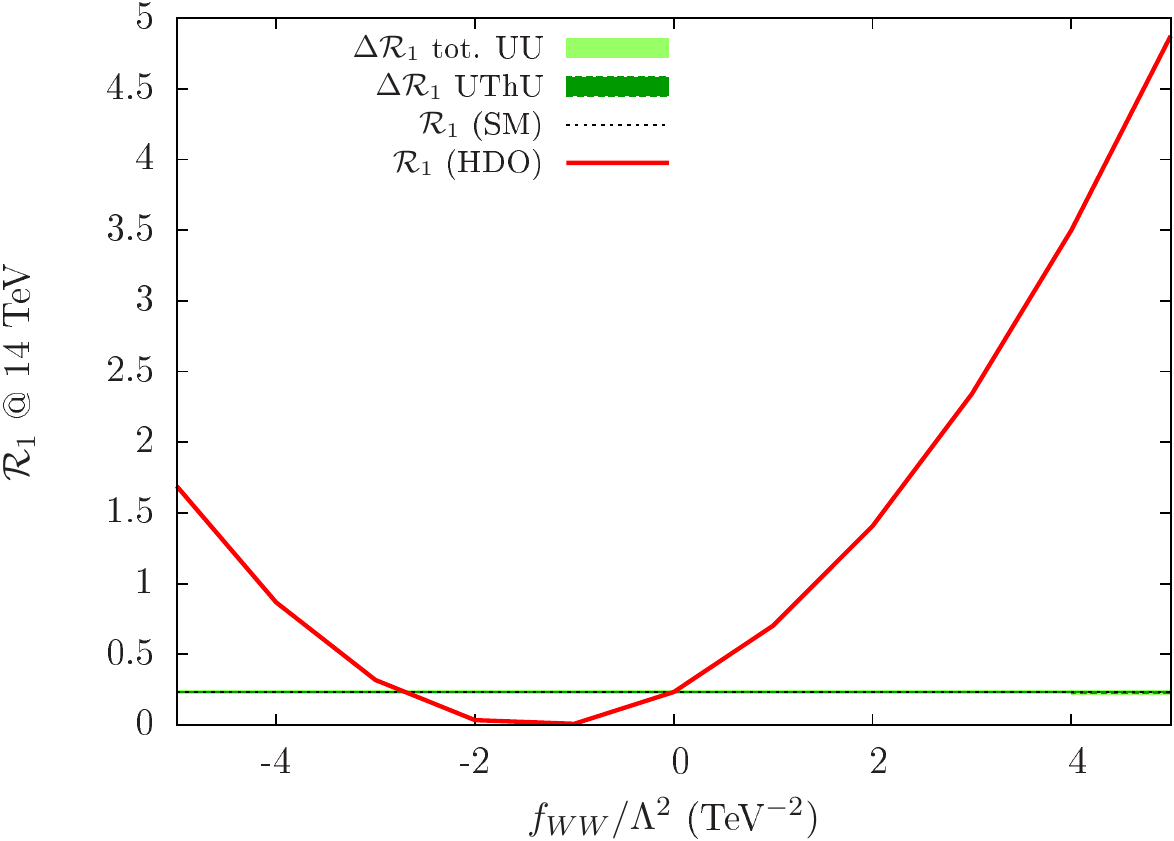}\label{fig:BRyywwfWW14}}~~~
\subfloat[]{\includegraphics[width=7cm,height=7cm]{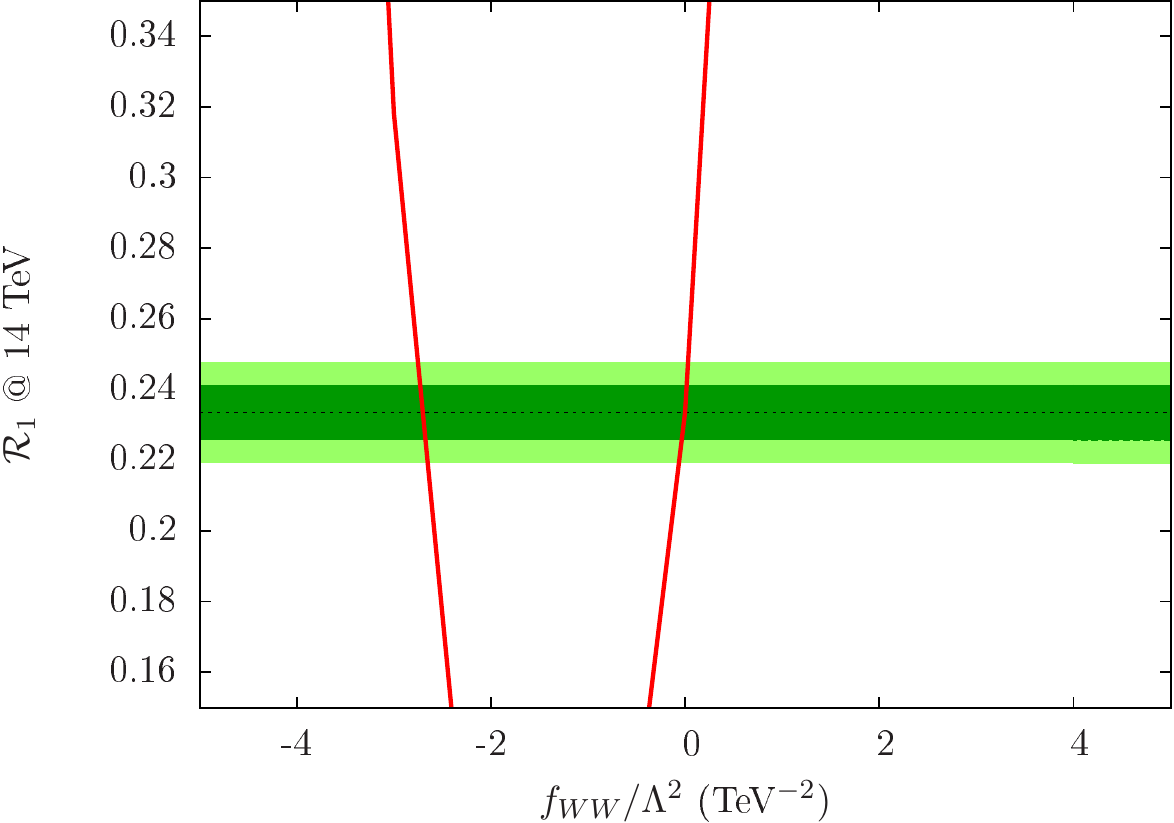}\label{fig:BRyywwfBB14}}
\caption{(a) $\mathcal{R}_1$ versus $f_{WW}/\Lambda^2$ (TeV$^{-2}$) and (b) same
plot in magnified scale. Plots (a) and (b) are identical for $f_{BB}/\Lambda^2$. The red line is the theoretical
  expectation in presence of HDOs. The inner band (dark green) shows the uncorrelated theoretical 
  uncertainty (UThU) and the outer band (light green) shows total
  uncorrelated uncertainty (UU) (uncorrelated theoretical + statistical + systematic) at
  14 TeV with 3000 fb$^{-1}$ integrated luminosity. The black dotted line is the corresponding central
  value. The uncertainty bands correspond to 68\% CL.}
\label{fig:R114}
\end{figure*}

 Next, we present a projected study of $\mathcal{R}_1$ for the 14 TeV run at 3000 fb$^{-1}$ of integrated luminosity.
 It should be noted here that the systematic uncertainties used here
 are for the 8 TeV run and we have assumed that they
 will not change significantly for the HL-LHC at 14 TeV. The inner bands,
 more clearly noticeable in Fig.~\ref{fig:BRyywwfBB14}, contain only the uncorrelated 
 theoretical
 errors, while the statistical and systematic errors are compounded in
 the outer bands. Clearly, the uncertainty gets reduced, as
 compared to $\mathcal{R}_1$ (@ 7 + 8 TeV), and we get an even smaller window
 around $f_{WW}$ and $f_{BB} \approx [-2.76, -2.65] \cup
 [-0.06,0.04]$ TeV$^{-2}$ as shown in Fig.~\ref{fig:R114}. The
 difference in this case is that the projected band is around the SM
 in contrast to what was shown for the 7+8 TeV case, where the ratio
 of the experimental signal strengths was treated as the reference.

\subsection{$\mathcal{R}_2$ @ 14 TeV}
\label{subsec:R214}

We now show the potential of $\mathcal{R}_2$ in deriving bounds on some of the
operator coefficients at 14 TeV. As is evident from Eq.~(\ref{eq:R2}),
this ratio has the capacity to probe $\mathcal{O}_{W}$ which cannot be
constrained from $\mathcal{R}_1$. On the other hand, the operator
$\mathcal{O}_{BB}$, though amenable to probe via $\mathcal{R}_1$, fails to show
any marked effect on $\mathcal{R}_2$ because $\tr{BR}_{H\to \gamma \gamma}$ 
gets cancelled in the ratio as defined by us. Also, $\mathcal{O}_{BB}$ does not
modify $\sigma_{\tr{WH}}$ but, $\mathcal{R}_2$ is however sensitive to the operator
$\mathcal{O}_{WW}$ as both $\sigma_{\tr{VBF}}$ and $\sigma_{\tr{WH}}$ are
sensitive to this.

By closely following the ATLAS
analyses in the context of high luminosity LHC run, we have used a trigger cut
of 50 GeV on jet $p_T$, instead of using $\eta$-dependent $p_T$ cut for jets as
used in Ref.~\cite{ATLAS-HL-LHC}. The reason is that, a flat cut on the $p_T$ will most certainly 
give us a less pessimistic number of final state events than that for the $\eta$ 
dependent $p_T$ cuts and performs as good as the $\eta$-dependent cut to suppress the background. So, we estimate a slightly larger number of events, {\it i.e.} we obtain a better 
efficiency to the cuts for the flat $p_T$ case as compared to what is predicted by ATLAS.
For the $WH$ production mode, we use a
matched sample with $WH+0,1,2$ jets with the $W$ decaying
leptonically. Finally we demand samples with a maximum of one jet in our analysis. In
selecting this $0+1$ jet sample, from a matched two jet sample, we
encounter another theoretical scale uncertainty as described in Ref.~\cite{Stewart:2011cf}.
We have estimated this uncertainty as follows:
\begin{equation}
\label{eq.stwtack}
\Delta^{th.}=\frac{\sigma(pp\to WH\, +\, \geq 2\,
    \textrm{jets})}{\sigma^{NNLO}(pp \to WH)}\Bigg\vert_{m_H} \times \Delta \sigma(pp\to WH\, +\, \geq 2\,
    \textrm{jets})(\mu_F,\mu_R),
\end{equation}
where $\Delta \sigma(pp\to WH\, +\, \geq 2\,
    \textrm{jets})$ is
the maximum deviation of the exclusive 2-jet cross-section computed at
$\mu_F=\mu_R=m_H$ from the ones computed by varying $\mu_F$ and
$\mu_R$ between $m_H/2$ and $2 m_H$.  
\begin{figure*}[ht!]
\centering
\subfloat[]{\includegraphics[width=7cm,height=7cm]{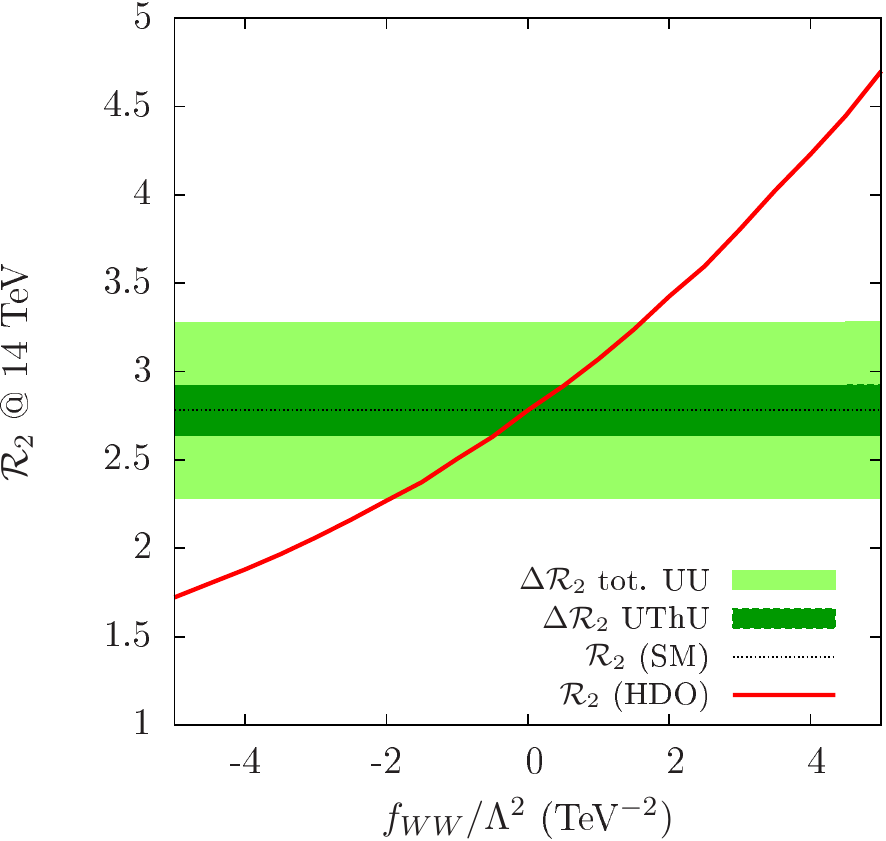}\label{fig:R2fWW}}~~~
\subfloat[]{\includegraphics[width=7cm,height=7cm]{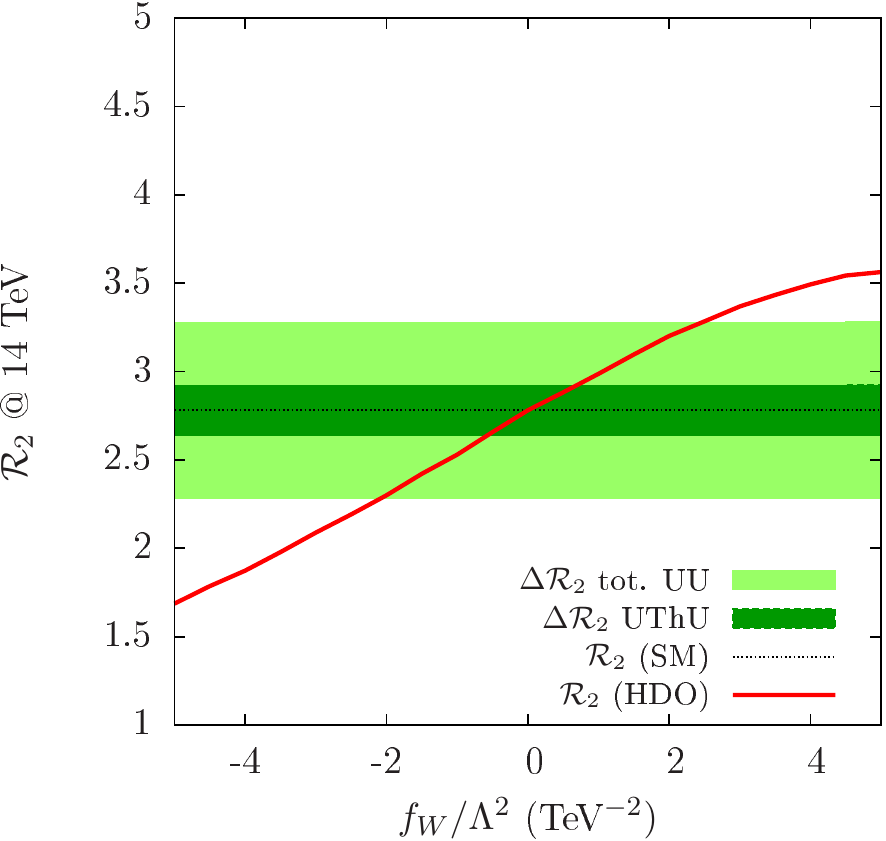}\label{fig:R2fW}}
\caption{The ratio $\mathcal{R}_2$ versus (a) $f_{WW}/\Lambda^2$ (TeV$^{-2}$), (b)
  $f_{W}/\Lambda^2$ (TeV$^{-2}$) for the 14 TeV analysis with 3000 fb$^{-1}$. The red line is the theoretical
  expectation in presence of HDOs. The inner band (dark green) shows the uncorrelated
  theoretical uncertainty due to PDF+$\alpha_s$, QCD-scale and
  $\Delta^{th.}$ which is defined in Eq.~(\ref{eq.stwtack}). The outer band (light green) shows the
  uncertainties due to the statistical, systematic compounded with the uncorrelated
  theoretical part. The black dotted line is the corresponding SM value. The uncertainty bands correspond to 68\% CL.}
\label{fig:R214}
\end{figure*}

In constructing $\mathcal{R}_2$, we include the modified
cut-efficiencies~\cite{Banerjee:2013apa,Gainer:2013rxa} for both the
$VBF$ and $WH$ channels. Even though we stick to small values of
$f_i$ where the modification in such efficiencies from the SM-values
are small, we still incorporate these to make the study more
rigorous. In computing the statistical uncertainties, we took the
relevant numbers from the 14 TeV projected study done by ATLAS (see Refs.~\cite{ATLAS-HL-LHC,ATLAS-HL-LHC-gaga}).
Besides, we also suggest tagging a single jet
for $VBF$, which reduces the statistical uncertainty by a factor of
$\sqrt{2}$~\cite{Kruse:2014pya}. The  $\sqrt{2}$ factor takes into account 
the number of events as well as the contamination due to $ggF$ as can be seen
on Table 1 in Ref.~\cite{Kruse:2014pya}. In Fig.~\ref{fig:R214}, we
present $\mathcal{R}_2$ as a function of the $f_{WW}$ and $f_W$ taken one at a
time for an integrated luminosity of $\mathcal{L}=3000$ fb$^{-1}$. The outer band (light green) shows the
  uncertainties due to the statistical, systematic compounded with the uncorrelated
  theoretical part. The central black dashed line
shows the SM expectation for $\mathcal{R}_2$. We can see in Fig.~\ref{fig:R214}
that very small values of HDO coefficients can be probed by measuring the 
observable $\mathcal{R}_2$. For $f_{WW}$, one can corner the allowed region to a small
window of $[-1.96,+1.62]$ and for $f_W$ the range would be $[-2.10,+2.50]$. Predicting the observability of such small values in the
parameter coefficients is definitely an improvement on existing knowledge.

\subsection{$\mathcal{R}_3$ @ 14 TeV}

\begin{figure*}[ht!]
\centering
\subfloat[]{\includegraphics[width=7cm,height=7cm]{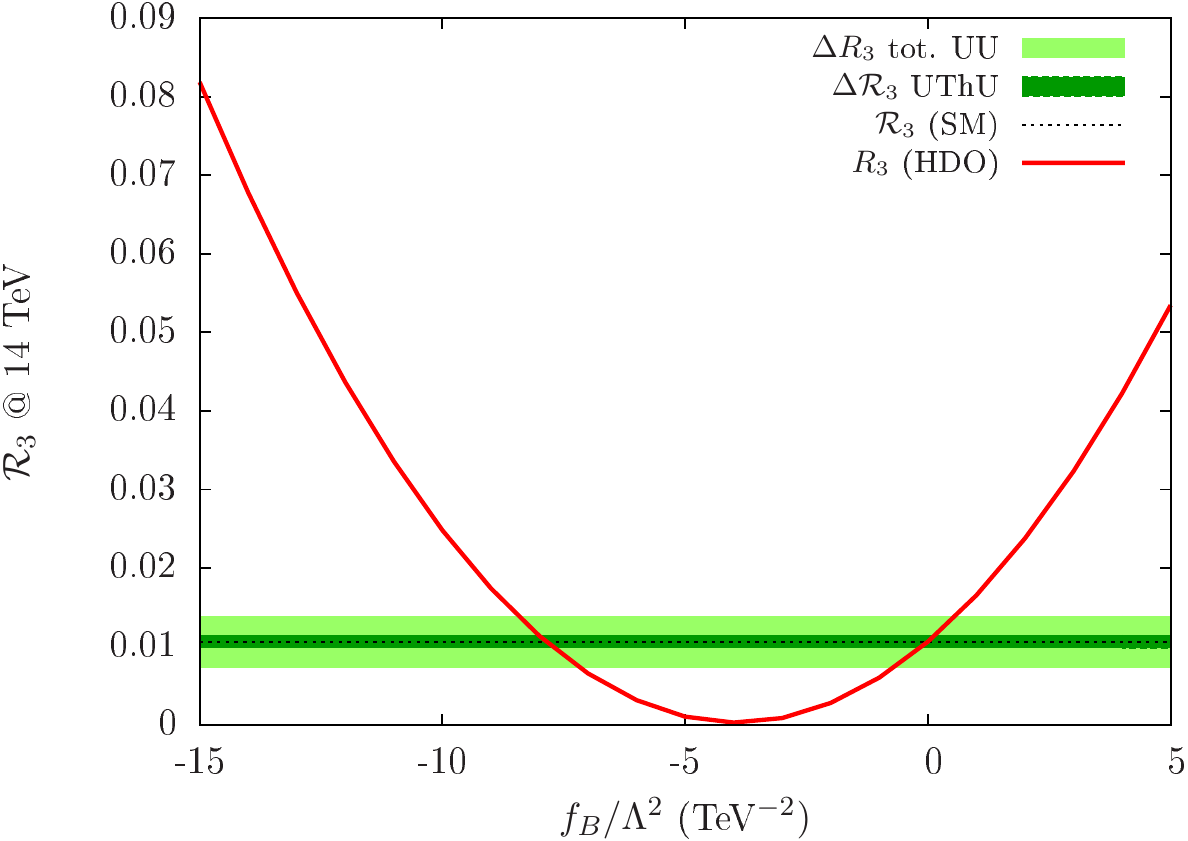}\label{fig:R3fB}}
\subfloat[]{\includegraphics[width=7cm,height=7cm]{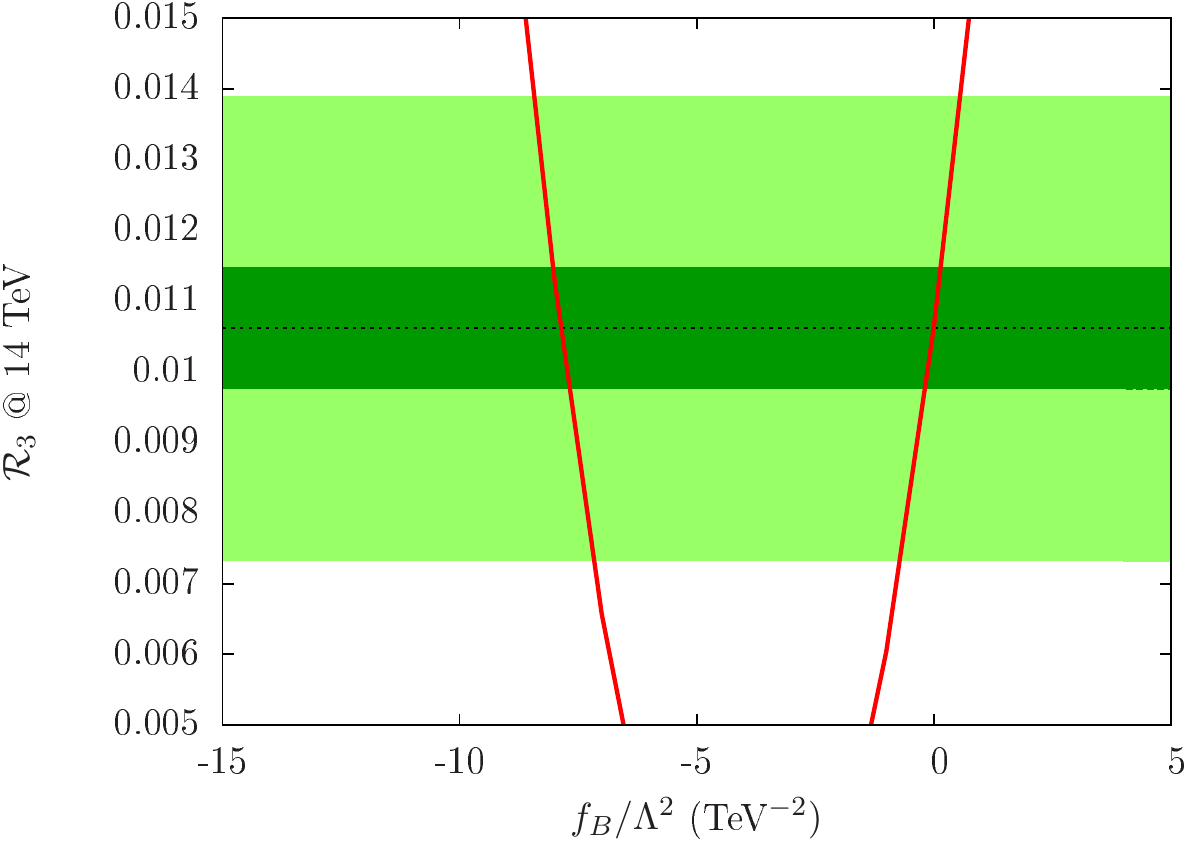}\label{fig:R3fBmag}}
\caption{The ratio $\mathcal{R}_3$ versus $f_{B}/\Lambda^2$ (TeV$^{-2}$) at 14 TeV
with 3000 fb$^{-1}$. The red line is
  the theoretical expectation in presence of HDOs. The inner band (dark green) shows the
  uncorrelated theoretical uncertainty (UThU) and the outer band (light green) shows
  the total uncorrelated uncertainty (UU) due to statistical, systematic and the
  uncorrelated theoretical part. These uncertainty bands are for $\mathcal{R}_3$
  at 14 TeV. The black dotted line is the corresponding SM value. The uncertainty bands correspond to 68\% CL.}
\label{fig:R314}
\end{figure*}
The operator $\mathcal{O}_{B}$ appears only in the $HZZ$ and
$HZ\gamma$ couplings, As seen in Fig.~\ref{fig:BRzz}, the sensitivity
of $\mathcal{O}_{B}$ is too low and hence $H\to ZZ^*$ will not give a
proper bound on $f_{B}/\Lambda^2$. Recent experiment by ATLAS (CMS)
puts bounds on the observed signal strength of $H\to Z\gamma$ at about 11
(9.5) times the SM expectation at $95$\% confidence level~\cite{Aad:2014fia,Chatrchyan:2013vaa}. Instead of
using these weak signal strengths, we perform an analogous projected
study of $\mathcal{R}_3$ at 14 TeV in the same spirit as $\mathcal{R}_1$ at 14 TeV.  
From
Fig.~\ref{fig:R314}, we find that the projected bounds on
$f_{B}/\Lambda^2$ is $[-8.44,-7.17]\cup [-0.72,+0.56]$. The region in
between is again inaccessible due to poor statistics, as in this
region, $\tr{BR}_{H\to Z \gamma}$ becomes insignificant, the reasons being
similar to those mentioned for $H\to \gamma \gamma$. The inner band (dark green)
includes the uncorrelated theoretical uncertainties due to the
partial decay widths of $H\to Z \gamma$ and $H \to WW^*$. The outer band (light
green), in addition to the theoretical uncertainties, contains the
statistical and systematic uncertainties. As discussed earlier, a few  
types of correlated systematic uncertainties related to the uncertainty in luminosity,
lepton identification and isolation etc. will get cancelled in the ratio $\mathcal{R}_3$.
On the other hand, photon identification, isolation etc. uncertainties will retain 
in the analysis.
\begin{table}
\centering
\begin{tabular}{|c|c|c|c|c|}
\hline 
Observable & $\mathcal{O}_{WW}$ & $\mathcal{O}_{BB}$ & $\mathcal{O}_{W}$ & $\mathcal{O}_{B}$ \\ 
\hline 
 & $[-3.32,-2.91]$ & $[-3.32,-2.91]$  & Not & Not \\ 
$\mathcal{R}_1$ @ 7+8 TeV & $\cup$ & $\cup$ & bounded & bounded\\
& $[+0.12,+0.57]$ & $[+0.12,+0.57]$ &  &  \\
\hline 
 & $[-2.76,-2.65]$ & $[-2.76,-2.65]$  & Not & Not \\ 
$\mathcal{R}_1$ @ 14 TeV & $\cup$ & $\cup$ & bounded & bounded \\
& $[-0.06,+0.04]$ & $[-0.06,+0.04]$ &  &  \\
\hline 
$\mathcal{R}_2$ @ 14 TeV & $[-1.96,+1.62]$ & Not & $[-2.10,+2.50]$ & Not  \\ 
& & bounded & & bounded \\
\hline 
& Not & Not & Not & $[-8.44,-7.17]$ \\
$\mathcal{R}_3$ @ 14 TeV  & used & used & used & $\cup$\\
&&&& $[-0.72,+0.56]$ \\ 
\hline 
\end{tabular}
\caption{We summarize our obtained allowed region of the coefficients of HDOs using the three observables. $\mathcal{R}_3$ is not used to constrain the operators $\mathcal{O}_{WW},\mathcal{O}_{BB}$ and $\mathcal{O}_W$ as has been discussed in Sec.~\ref{subsec:R3}.} 
\label{tab:limit}
\end{table}
In Table~\ref{tab:limit}, we summarize our obtained region of the parameter space allowed using
three ratios, $\mathcal{R}_1$, $\mathcal{R}_2$ and $\mathcal{R}_3$. We present $\mathcal{R}_1$
using combined ATLAS+CMS data for 7+8 TeV run. We also present a projected study for all three
observables at 14 TeV with an integrated luminosity of 3000 fb$^{-1}$. The allowed regions on 
$f_{WW}$ and $f_{BB}$ shrink at the 14 TeV 3000 fb$^{-1}$ run as compared to the current data. Using
the ratio, $\mathcal{R}_2$ one can also put bounds on $f_{WW}$ and $f_{W}$. As mentioned 
earlier, there is a `special' region of parameter space where $\mathcal{R}_1$ mimics
the SM expectation, therefore, $\mathcal{R}_2$ can also be used to infer the presence of $\mathcal{O}_{WW}$
with `special' values of coefficient $f_{WW}$. The operator $\mathcal{O}_B$ does not show any 
appreciable sensitivity in any production of Higgs or its decay except in the $\tr{BR}_{H\to Z\gamma}$.
Therefore, the ratio $\mathcal{R}_3$ is constructed to constrain $f_{B}$ by a significant amount as
evident from Table~\ref{tab:limit}.
\section{Summary and conclusions}
\label{summary}

We have investigated how well one can constrain dimension-6 gauge-invariant
operators inducing anomalous $HVV$ interactions. Probing the gauge invariant operators
individually, we feel, are important, since they can point at any new physics above
the electroweak symmetry breaking scale.  While the operators contributing to
$H\to \gamma\gamma$ are subjected to the hitherto strongest limits using the
(7+8) TeV data, the remaining ones are relatively loosely constrained, in spite of the
bounds coming from precision electroweak observables. At any rate, it is necessary to
reduce uncertainties as much as possible, since any realistically conceived new physics
is likely to generate such operators with coefficients no greater than
$\approx \mathcal{O}(1)$ TeV$^{-2}$. We show that a good opportunity to probe them at this level,
and improve spectacularly over the existing constraints, arises if event ratios in various
channels are carefully studied. These include both ratios of events in different final states
with the same Higgs production channel and those where a Higgs produced by different production modes ends up decaying into the same final state. While a majority of the theoretical uncertainties
cancel in the former category, the latter  allow us to probe those cases where
some dimension-6 operators shift the rates in the  numerator and the denominator
in opposite directions. We find that, after a thorough consideration of all
uncertainties, all the couplings can be pinned down to intervals of width
$\approx \mathcal{O}(1)$ TeV$^{-2}$ on using 3000 fb$^{-1}$ of integrated luminosity at
14 TeV. Even with 300  fb$^{-1}$, the improvement over existing constraints
is clearly expected, and the results are more uncertainty-free than in any
other hitherto applied method. However, we must mention here that this approach should be 
complemented with the study of differential distributions which is not within the scope of this paper.

\section*{Acknowledgements}
The work of S.B., T.M. and B. Mukhopadhyaya was partially 
supported by funding available from the Department of Atomic Energy, Government
of India for the Regional Centre for Accelerator-based Particle
Physics (RECAPP), Harish-Chandra Research Institute. B. Mellado acknowledges
the hospitality of RECAPP, Harish-Chandra Research Institute, during
the collaboration.

\bibliographystyle{JHEP}
\bibliography{HDopsLHC}{}

\providecommand{\href}[2]{#2}\begingroup\raggedright\begin{thebibliography}{10}

\bibitem{Aad:2012tfa}
{\bf ATLAS} Collaboration, G.~Aad et~al., {\it {Observation of a new particle
  in the search for the Standard Model Higgs boson with the ATLAS detector at
  the LHC}},  {\em Phys.Lett.} {\bf B716} (2012) 1--29,
  [\href{http://arxiv.org/abs/1207.7214}{{\tt arXiv:1207.7214}}].

\bibitem{Chatrchyan:2012ufa}
{\bf CMS} Collaboration, S.~Chatrchyan et~al., {\it {Observation of a new boson
  at a mass of 125 GeV with the CMS experiment at the LHC}},  {\em Phys.Lett.}
  {\bf B716} (2012) 30--61, [\href{http://arxiv.org/abs/1207.7235}{{\tt
  arXiv:1207.7235}}].

\bibitem{Masso:2012eq}
E.~Masso and V.~Sanz, {\it {Limits on anomalous couplings of the Higgs boson
  to electroweak gauge bosons from LEP and the LHC}},  {\em Phys.Rev.} {\bf
  D87} (2013), no.~3 033001, [\href{http://arxiv.org/abs/1211.1320}{{\tt
  arXiv:1211.1320}}].

\bibitem{Corbett:2012ja}
T.~Corbett, O.~Eboli, J.~Gonzalez-Fraile, and M.~Gonzalez-Garcia, {\it {Robust
  Determination of the Higgs Couplings: Power to the Data}},  {\em Phys.Rev.}
  {\bf D87} (2013) 015022, [\href{http://arxiv.org/abs/1211.4580}{{\tt
  arXiv:1211.4580}}].

\bibitem{Falkowski:2013dza}
A.~Falkowski, F.~Riva, and A.~Urbano, {\it {Higgs at last}},  {\em JHEP} {\bf
  1311} (2013) 111, [\href{http://arxiv.org/abs/1303.1812}{{\tt
  arXiv:1303.1812}}].

\bibitem{Corbett:2013pja}
T.~Corbett, O.~Éboli, J.~Gonzalez-Fraile, and M.~Gonzalez-Garcia, {\it
  {Determining Triple Gauge Boson Couplings from Higgs Data}},  {\em
  Phys.Rev.Lett.} {\bf 111} (2013) 011801,
  [\href{http://arxiv.org/abs/1304.1151}{{\tt arXiv:1304.1151}}].

\bibitem{Dumont:2013wma}
B.~Dumont, S.~Fichet, and G.~von Gersdorff, {\it {A Bayesian view of the Higgs
  sector with higher dimensional operators}},  {\em JHEP} {\bf 1307} (2013)
  065, [\href{http://arxiv.org/abs/1304.3369}{{\tt arXiv:1304.3369}}].

\bibitem{Banerjee:2012xc}
S.~Banerjee, S.~Mukhopadhyay, and B.~Mukhopadhyaya, {\it {New Higgs
  interactions and recent data from the LHC and the Tevatron}},  {\em JHEP}
  {\bf 1210} (2012) 062, [\href{http://arxiv.org/abs/1207.3588}{{\tt
  arXiv:1207.3588}}].

\bibitem{Gainer:2013rxa}
J.~S. Gainer, J.~Lykken, K.~T. Matchev, S.~Mrenna, and M.~Park, {\it
  {Geolocating the Higgs Boson Candidate at the LHC}},  {\em Phys.Rev.Lett.}
  {\bf 111} (2013) 041801, [\href{http://arxiv.org/abs/1304.4936}{{\tt
  arXiv:1304.4936}}].

\bibitem{Corbett:2013hia}
T.~Corbett, O.~Éboli, J.~Gonzalez-Fraile, and M.~Gonzalez-Garcia, {\it {Robust
  determination of the scalar boson couplings}},
  \href{http://arxiv.org/abs/1306.0006}{{\tt arXiv:1306.0006}}.

\bibitem{Elias-Miro:2013mua}
J.~Elias-Miro, J.~Espinosa, E.~Masso, and A.~Pomarol, {\it {Higgs windows to
  new physics through d=6 operators: constraints and one-loop anomalous
  dimensions}},  {\em JHEP} {\bf 1311} (2013) 066,
  [\href{http://arxiv.org/abs/1308.1879}{{\tt arXiv:1308.1879}}].

\bibitem{Pomarol:2013zra}
A.~Pomarol and F.~Riva, {\it {Towards the Ultimate SM Fit to Close in on Higgs
  Physics}},  {\em JHEP} {\bf 1401} (2014) 151,
  [\href{http://arxiv.org/abs/1308.2803}{{\tt arXiv:1308.2803}}].

\bibitem{Einhorn:2013tja}
M.~B. Einhorn and J.~Wudka, {\it {Higgs-Boson Couplings Beyond the Standard
  Model}},  {\em Nucl.Phys.} {\bf B877} (2013) 792--806,
  [\href{http://arxiv.org/abs/1308.2255}{{\tt arXiv:1308.2255}}].

\bibitem{Banerjee:2013apa}
S.~Banerjee, S.~Mukhopadhyay, and B.~Mukhopadhyaya, {\it {Higher dimensional
  operators and the LHC Higgs data: The role of modified kinematics}},  {\em
  Phys.Rev.} {\bf D89} (2014), no.~5 053010,
  [\href{http://arxiv.org/abs/1308.4860}{{\tt arXiv:1308.4860}}].

\bibitem{Willenbrock:2014bja}
S.~Willenbrock and C.~Zhang, {\it {Effective Field Theory Beyond the Standard
  Model}},  {\em Ann.Rev.Nucl.Part.Sci.} {\bf 64} (2014) 83--100,
  [\href{http://arxiv.org/abs/1401.0470}{{\tt arXiv:1401.0470}}].

\bibitem{Ellis:2014dva}
J.~Ellis, V.~Sanz, and T.~You, {\it {Complete Higgs Sector Constraints on
  Dimension-6 Operators}},  {\em JHEP} {\bf 1407} (2014) 036,
  [\href{http://arxiv.org/abs/1404.3667}{{\tt arXiv:1404.3667}}].

\bibitem{Belusca-Maito:2014dpa}
H.~Belusca-Maito, {\it {Effective Higgs Lagrangian and Constraints on Higgs
  Couplings}},  \href{http://arxiv.org/abs/1404.5343}{{\tt arXiv:1404.5343}}.

\bibitem{Gupta:2014rxa}
R.~S. Gupta, A.~Pomarol, and F.~Riva, {\it {BSM Primary Effects}},  {\em
  Phys.Rev.} {\bf D91} (2015), no.~3 035001,
  [\href{http://arxiv.org/abs/1405.0181}{{\tt arXiv:1405.0181}}].

\bibitem{Masso:2014xra}
E.~Masso, {\it {An Effective Guide to Beyond the Standard Model Physics}},
  {\em JHEP} {\bf 1410} (2014) 128, [\href{http://arxiv.org/abs/1406.6376}{{\tt
  arXiv:1406.6376}}].

\bibitem{Biekoetter:2014jwa}
A.~Biekötter, A.~Knochel, M.~Krämer, D.~Liu, and F.~Riva, {\it {Vices and
  virtues of Higgs effective field theories at large energy}},  {\em Phys.Rev.}
  {\bf D91} (2015) 055029, [\href{http://arxiv.org/abs/1406.7320}{{\tt
  arXiv:1406.7320}}].

\bibitem{Englert:2014cva}
C.~Englert and M.~Spannowsky, {\it {Effective Theories and Measurements at
  Colliders}},  {\em Phys.Lett.} {\bf B740} (2015) 8--15,
  [\href{http://arxiv.org/abs/1408.5147}{{\tt arXiv:1408.5147}}].

\bibitem{Ellis:2014jta}
J.~Ellis, V.~Sanz, and T.~You, {\it {The Effective Standard Model after LHC Run
  I}},  {\em JHEP} {\bf 1503} (2015) 157,
  [\href{http://arxiv.org/abs/1410.7703}{{\tt arXiv:1410.7703}}].

\bibitem{Edezhath:2015lga}
R.~Edezhath, {\it {Dimension-6 Operator Constraints from Boosted VBF Higgs}},
  \href{http://arxiv.org/abs/1501.00992}{{\tt arXiv:1501.00992}}.

\bibitem{Gorbahn:2015gxa}
M.~Gorbahn, J.~M. No, and V.~Sanz, {\it {Benchmarks for Higgs Effective Theory:
  Extended Higgs Sectors}},  \href{http://arxiv.org/abs/1502.07352}{{\tt
  arXiv:1502.07352}}.

\bibitem{Han:2004az}
Z.~Han and W.~Skiba, {\it {Effective theory analysis of precision electroweak
  data}},  {\em Phys.Rev.} {\bf D71} (2005) 075009,
  [\href{http://arxiv.org/abs/hep-ph/0412166}{{\tt hep-ph/0412166}}].

\bibitem{Ciuchini:2013pca}
M.~Ciuchini, E.~Franco, S.~Mishima, and L.~Silvestrini, {\it {Electroweak
  Precision Observables, New Physics and the Nature of a 126 GeV Higgs Boson}},
   {\em JHEP} {\bf 1308} (2013) 106,
  [\href{http://arxiv.org/abs/1306.4644}{{\tt arXiv:1306.4644}}].

\bibitem{Blas:2013ana}
J.~de~Blas, {\it {Electroweak limits on physics beyond the Standard Model}},
  {\em EPJ Web Conf.} {\bf 60} (2013) 19008,
  [\href{http://arxiv.org/abs/1307.6173}{{\tt arXiv:1307.6173}}].

\bibitem{Chen:2013kfa}
C.-Y. Chen, S.~Dawson, and C.~Zhang, {\it {Electroweak Effective Operators and
  Higgs Physics}},  {\em Phys.Rev.} {\bf D89} (2014), no.~1 015016,
  [\href{http://arxiv.org/abs/1311.3107}{{\tt arXiv:1311.3107}}].

\bibitem{Alonso:2013hga}
R.~Alonso, E.~E. Jenkins, A.~V. Manohar, and M.~Trott, {\it {Renormalization
  Group Evolution of the Standard Model Dimension Six Operators III: Gauge
  Coupling Dependence and Phenomenology}},  {\em JHEP} {\bf 1404} (2014) 159,
  [\href{http://arxiv.org/abs/1312.2014}{{\tt arXiv:1312.2014}}].

\bibitem{Englert:2014uua}
C.~Englert, A.~Freitas, M.~Mühlleitner, T.~Plehn, M.~Rauch, et~al., {\it
  {Precision Measurements of Higgs Couplings: Implications for New Physics
  Scales}},  {\em J.Phys.} {\bf G41} (2014) 113001,
  [\href{http://arxiv.org/abs/1403.7191}{{\tt arXiv:1403.7191}}].

\bibitem{Trott:2014dma}
M.~Trott, {\it {On the consistent use of Constructed Observables}},  {\em JHEP}
  {\bf 1502} (2015) 046, [\href{http://arxiv.org/abs/1409.7605}{{\tt
  arXiv:1409.7605}}].

\bibitem{Falkowski:2014tna}
A.~Falkowski and F.~Riva, {\it {Model-independent precision constraints on
  dimension-6 operators}},  {\em JHEP} {\bf 1502} (2015) 039,
  [\href{http://arxiv.org/abs/1411.0669}{{\tt arXiv:1411.0669}}].

\bibitem{Henning:2014wua}
B.~Henning, X.~Lu, and H.~Murayama, {\it {How to use the Standard Model
  effective field theory}},  \href{http://arxiv.org/abs/1412.1837}{{\tt
  arXiv:1412.1837}}.

\bibitem{deBlas:2014mba}
J.~de~Blas, M.~Chala, M.~Perez-Victoria, and J.~Santiago, {\it {Observable
  Effects of General New Scalar Particles}},
  \href{http://arxiv.org/abs/1412.8480}{{\tt arXiv:1412.8480}}.

\bibitem{Berthier:2015oma}
L.~Berthier and M.~Trott, {\it {Towards consistent Electroweak Precision Data
  constraints in the SMEFT}},  \href{http://arxiv.org/abs/1502.02570}{{\tt
  arXiv:1502.02570}}.

\bibitem{Efrati:2015eaa}
A.~Efrati, A.~Falkowski, and Y.~Soreq, {\it {Electroweak constraints on
  flavorful effective theories}},  \href{http://arxiv.org/abs/1503.07872}{{\tt
  arXiv:1503.07872}}.

\bibitem{Bhattacherjee:2015xra}
B.~Bhattacherjee, T.~Modak, S.~K. Patra, and R.~Sinha, {\it {Probing Higgs
  couplings at LHC and beyond}},  \href{http://arxiv.org/abs/1503.08924}{{\tt
  arXiv:1503.08924}}.

\bibitem{Khachatryan:2014kca}
{\bf CMS} Collaboration, V.~Khachatryan et~al., {\it {Constraints on the
  spin-parity and anomalous HVV couplings of the Higgs boson in proton
  collisions at 7 and 8 TeV}},  \href{http://arxiv.org/abs/1411.3441}{{\tt
  arXiv:1411.3441}}.

\bibitem{Amar:2014fpa}
G.~Amar, S.~Banerjee, S.~von Buddenbrock, A.~S. Cornell, T.~Mandal, et~al.,
  {\it {Exploration of the tensor structure of the Higgs boson coupling to weak
  bosons in e$^{+}$ e$^{−}$ collisions}},  {\em JHEP} {\bf 1502} (2015) 128,
  [\href{http://arxiv.org/abs/1405.3957}{{\tt arXiv:1405.3957}}].

\bibitem{Kumar:2014zra}
S.~Kumar and P.~Poulose, {\it {Influence of anomalous VVH and VVHH on
  determination of Higgs self couplings at ILC}},
  \href{http://arxiv.org/abs/1408.3563}{{\tt arXiv:1408.3563}}.

\bibitem{Craig:2014una}
N.~Craig, M.~Farina, M.~McCullough, and M.~Perelstein, {\it {Precision
  Higgsstrahlung as a Probe of New Physics}},  {\em JHEP} {\bf 1503} (2015)
  146, [\href{http://arxiv.org/abs/1411.0676}{{\tt arXiv:1411.0676}}].

\bibitem{Beneke:2014vqa}
M.~Beneke, D.~Boito, and Y.-M. Wang, {\it {Signatures of anomalous Higgs
  couplings in angular asymmetries of $H \to Z\ell^+\ell^-$ and $e^+e^- \to HZ$}},
  \href{http://arxiv.org/abs/1411.3942}{{\tt arXiv:1411.3942}}.

\bibitem{Kumar:2015eea}
S.~Kumar, P.~Poulose, and S.~Sahoo, {\it {Study of Higgs-gauge boson anomalous
  couplings through $e^-e^+ \to W^-W^+H$ at ILC}},
  \href{http://arxiv.org/abs/1501.03283}{{\tt arXiv:1501.03283}}.

\bibitem{Ren:2015uka}
H.-Y. Ren, {\it {New Physics Searches with Higgs-photon associated production
  at the Higgs Factory}},  \href{http://arxiv.org/abs/1503.08307}{{\tt
  arXiv:1503.08307}}.

\bibitem{Plehn:2001nj}
T.~Plehn, D.~L. Rainwater, and D.~Zeppenfeld, {\it {Determining the structure
  of Higgs couplings at the LHC}},  {\em Phys.Rev.Lett.} {\bf 88} (2002)
  051801, [\href{http://arxiv.org/abs/hep-ph/0105325}{{\tt hep-ph/0105325}}].

\bibitem{Bernaciak:2012nh}
C.~Bernaciak, M.~S.~A. Buschmann, A.~Butter, and T.~Plehn, {\it {Fox-Wolfram
  Moments in Higgs Physics}},  {\em Phys.Rev.} {\bf D87} (2013) 073014,
  [\href{http://arxiv.org/abs/1212.4436}{{\tt arXiv:1212.4436}}].

\bibitem{Bernaciak:2013dwa}
C.~Bernaciak, B.~Mellado, T.~Plehn, P.~Schichtel, and X.~Ruan, {\it {Improving
  Higgs plus Jets analyses through Fox--Wolfram Moments}},  {\em Phys.Rev.}
  {\bf D89} (2014), no.~5 053006, [\href{http://arxiv.org/abs/1311.5891}{{\tt
  arXiv:1311.5891}}].

\bibitem{Biswal:2012mp}
S.~S. Biswal, R.~M. Godbole, B.~Mellado, and S.~Raychaudhuri, {\it {Azimuthal
  Angle Probe of Anomalous $HWW$ Couplings at a High Energy $ep$ Collider}},
  {\em Phys.Rev.Lett.} {\bf 109} (2012) 261801,
  [\href{http://arxiv.org/abs/1203.6285}{{\tt arXiv:1203.6285}}].

\bibitem{Djouadi:2013yb}
A.~Djouadi, R.~Godbole, B.~Mellado, and K.~Mohan, {\it {Probing the spin-parity
  of the Higgs boson via jet kinematics in vector boson fusion}},  {\em
  Phys.Lett.} {\bf B723} (2013) 307--313,
  [\href{http://arxiv.org/abs/1301.4965}{{\tt arXiv:1301.4965}}].

\bibitem{Djouadi:2012rh}
A.~Djouadi, {\it {Precision Higgs coupling measurements at the LHC through
  ratios of production cross sections}},  {\em Eur.Phys.J.} {\bf C73} (2013)
  2498, [\href{http://arxiv.org/abs/1208.3436}{{\tt arXiv:1208.3436}}].

\bibitem{Djouadi:2013qya}
A.~Djouadi and G.~Moreau, {\it {The couplings of the Higgs boson and its CP
  properties from fits of the signal strengths and their ratios at the 7+8 TeV
  LHC}},  {\em Eur.Phys.J.} {\bf C73} (2013), no.~9 2512,
  [\href{http://arxiv.org/abs/1303.6591}{{\tt arXiv:1303.6591}}].

\bibitem{Buchmuller:1985jz}
W.~Buchmuller and D.~Wyler, {\it {Effective Lagrangian Analysis of New
  Interactions and Flavor Conservation}},  {\em Nucl.Phys.} {\bf B268} (1986)
  621--653.

\bibitem{Hagiwara:1993qt}
K.~Hagiwara, R.~Szalapski, and D.~Zeppenfeld, {\it {Anomalous Higgs boson
  production and decay}},  {\em Phys.Lett.} {\bf B318} (1993) 155--162,
  [\href{http://arxiv.org/abs/hep-ph/9308347}{{\tt hep-ph/9308347}}].

\bibitem{GonzalezGarcia:1999fq}
M.~Gonzalez-Garcia, {\it {Anomalous Higgs couplings}},  {\em Int.J.Mod.Phys.}
  {\bf A14} (1999) 3121--3156, [\href{http://arxiv.org/abs/hep-ph/9902321}{{\tt
  hep-ph/9902321}}].

\bibitem{Grzadkowski:2010es}
B.~Grzadkowski, M.~Iskrzynski, M.~Misiak, and J.~Rosiek, {\it {Dimension-Six
  Terms in the Standard Model Lagrangian}},  {\em JHEP} {\bf 1010} (2010) 085,
  [\href{http://arxiv.org/abs/1008.4884}{{\tt arXiv:1008.4884}}].

\bibitem{Einhorn:2013kja}
M.~B. Einhorn and J.~Wudka, {\it {The Bases of Effective Field Theories}},
  {\em Nucl.Phys.} {\bf B876} (2013) 556--574,
  [\href{http://arxiv.org/abs/1307.0478}{{\tt arXiv:1307.0478}}].

\bibitem{Aad:2014fia}
{\bf ATLAS} Collaboration, G.~Aad et~al., {\it {Search for Higgs boson decays
  to a photon and a Z boson in pp collisions at $\sqrt{s}$=7 and 8 TeV with the
  ATLAS detector}},  {\em Phys.Lett.} {\bf B732} (2014) 8--27,
  [\href{http://arxiv.org/abs/1402.3051}{{\tt arXiv:1402.3051}}].

\bibitem{Chatrchyan:2013vaa}
{\bf CMS} Collaboration, S.~Chatrchyan et~al., {\it {Search for a Higgs boson
  decaying into a Z and a photon in pp collisions at $\sqrt{s}$ = 7 and 8
  TeV}},  {\em Phys.Lett.} {\bf B726} (2013) 587--609,
  [\href{http://arxiv.org/abs/1307.5515}{{\tt arXiv:1307.5515}}].

\bibitem{Heinemeyer:2013tqa}
{\bf LHC Higgs Cross Section Working Group} Collaboration, S.~Heinemeyer
  et~al., {\it {Handbook of LHC Higgs Cross Sections: 3. Higgs Properties}},
  \href{http://arxiv.org/abs/1307.1347}{{\tt arXiv:1307.1347}}.

\bibitem{Khachatryan:2014jba}
{\bf CMS} Collaboration, V.~Khachatryan et~al., {\it {Precise determination of
  the mass of the Higgs boson and tests of compatibility of its couplings with
  the standard model predictions using proton collisions at 7 and 8 TeV}},
  \href{http://arxiv.org/abs/1412.8662}{{\tt arXiv:1412.8662}}.

\bibitem{ATLAS-CONF-2015-007}
{\it {Measurements of the Higgs boson production and decay rates and coupling
  strengths using pp collision data at √s = 7 and 8 TeV in the ATLAS
  experiment}},  Tech. Rep. ATLAS-CONF-2015-007, CERN, Geneva, Mar, 2015.

\bibitem{Alloul:2013bka}
A.~Alloul, N.~D. Christensen, C.~Degrande, C.~Duhr, and B.~Fuks, {\it
  {FeynRules 2.0 - A complete toolbox for tree-level phenomenology}},  {\em
  Comput.Phys.Commun.} {\bf 185} (2014) 2250--2300,
  [\href{http://arxiv.org/abs/1310.1921}{{\tt arXiv:1310.1921}}].

\bibitem{Degrande:2011ua}
C.~Degrande, C.~Duhr, B.~Fuks, D.~Grellscheid, O.~Mattelaer, et~al., {\it {UFO
  - The Universal FeynRules Output}},  {\em Comput.Phys.Commun.} {\bf 183}
  (2012) 1201--1214, [\href{http://arxiv.org/abs/1108.2040}{{\tt
  arXiv:1108.2040}}].

\bibitem{Alwall:2014hca}
J.~Alwall, R.~Frederix, S.~Frixione, V.~Hirschi, F.~Maltoni, et~al., {\it {The
  automated computation of tree-level and next-to-leading order differential
  cross sections, and their matching to parton shower simulations}},  {\em
  JHEP} {\bf 1407} (2014) 079, [\href{http://arxiv.org/abs/1405.0301}{{\tt
  arXiv:1405.0301}}].

\bibitem{Sjostrand:2006za}
T.~Sjostrand, S.~Mrenna, and P.~Z. Skands, {\it {PYTHIA 6.4 Physics and
  Manual}},  {\em JHEP} {\bf 0605} (2006) 026,
  [\href{http://arxiv.org/abs/hep-ph/0603175}{{\tt hep-ph/0603175}}].

\bibitem{deFavereau:2013fsa}
{\bf DELPHES 3} Collaboration, J.~de~Favereau et~al., {\it {DELPHES 3, A
  modular framework for fast simulation of a generic collider experiment}},
  {\em JHEP} {\bf 1402} (2014) 057, [\href{http://arxiv.org/abs/1307.6346}{{\tt
  arXiv:1307.6346}}].

\bibitem{twiki}
``L\uppercase{H}\uppercase{C} \uppercase{H}iggs \uppercase{C}ross
  \uppercase{S}ection \uppercase{W}orking \uppercase{G}roup.''
  \url{https://twiki.cern.ch/twiki/bin/view/LHCPhysics/CrossSections}.

\bibitem{ATLAS-HL-LHC}
{\bf ATLAS} Collaboration, G.~Aad et~al., ``{Projections for measurements of
  Higgs boson cross sections, branching ratios and coupling parameters with the
  ATLAS detector at a HL-LHC}.''
  \url{https://atlas.web.cern.ch/Atlas/GROUPS/PHYSICS/PUBNOTES/ATL-PHYS-PUB-2013-014/},
  2013.

\bibitem{ATLAS-HL-LHC-gaga}
{\bf ATLAS} Collaboration, G.~Aad et~al., ``{HL-LHC projections for signal and
  background yield measurements of the $H \to \gamma \gamma$ when the Higgs
  boson is produced in association with $t$ quarks, $W$ or $Z$ bosons}.''
  \url{http://atlas.web.cern.ch/Atlas/GROUPS/PHYSICS/PUBNOTES/ATL-PHYS-PUB-2014-012/},
  2014.

\bibitem{Aad:2014eha}
{\bf ATLAS} Collaboration, G.~Aad et~al., {\it {Measurement of Higgs boson
  production in the diphoton decay channel in pp collisions at center-of-mass
  energies of 7 and 8 TeV with the ATLAS detector}},  {\em Phys.Rev.} {\bf D90}
  (2014), no.~11 112015, [\href{http://arxiv.org/abs/1408.7084}{{\tt
  arXiv:1408.7084}}].

\bibitem{ATLAS:2014aga}
{\bf ATLAS} Collaboration, G.~Aad et~al., {\it {Observation and measurement of
  Higgs boson decays to $WW^{\ast}$ with the ATLAS detector}},
  \href{http://arxiv.org/abs/1412.2641}{{\tt arXiv:1412.2641}}.

\bibitem{Khachatryan:2014ira}
{\bf CMS} Collaboration, V.~Khachatryan et~al., {\it {Observation of the
  diphoton decay of the Higgs boson and measurement of its properties}},  {\em
  Eur.Phys.J.} {\bf C74} (2014), no.~10 3076,
  [\href{http://arxiv.org/abs/1407.0558}{{\tt arXiv:1407.0558}}].

\bibitem{CMS:2014ega}
{\bf CMS} Collaboration, S.~Chatrchyan et~al., ``{Precise determination of the
  mass of the Higgs boson and studies of the compatibility of its couplings
  with the standard model}.'' \url{http://cds.cern.ch/record/1728249}, 2014.

\bibitem{Stewart:2011cf}
I.~W. Stewart and F.~J. Tackmann, {\it {Theory Uncertainties for Higgs and
  Other Searches Using Jet Bins}},  {\em Phys.Rev.} {\bf D85} (2012) 034011,
  [\href{http://arxiv.org/abs/1107.2117}{{\tt arXiv:1107.2117}}].

\bibitem{Kruse:2014pya}
A.~Kruse, A.~S. Cornell, M.~Kumar, B.~Mellado, and X.~Ruan, {\it {Probing the
  Higgs boson via vector boson fusion with single jet tagging at the LHC}},
  {\em Phys.Rev.} {\bf D91} (2015), no.~5 053009,
  [\href{http://arxiv.org/abs/1412.4710}{{\tt arXiv:1412.4710}}].

\end{thebibliography}\endgroup

\end{document}